\documentclass[11pt]{article}
\usepackage{comment}
\usepackage[hidelinks]{hyperref}
\usepackage{enumitem}
\usepackage{amsmath,amssymb,epsf,cite,graphicx,subfigure}
\usepackage{amsmath,amsthm, tikzsymbols, dsfont, amsfonts}
\usepackage{algorithm}
\usepackage{algorithmic}
\usepackage{forloop}
\usepackage{comment}
\usepackage{xcolor}
\usepackage{graphicx}
\usepackage{dsfont}
\usepackage{soul}
\newcommand{\R}{\mathbb{R}}

\allowdisplaybreaks

\newcommand{\M}{\mathcal{M}}
\DeclareMathOperator*{\argmin}{arg\,min}
\DeclareMathOperator*{\E}{\mathbb{E}}
\newcommand{\Grad}{\nabla}
\numberwithin{equation}{section}

\definecolor{airforceblue}{rgb}{0.36, 0.54, 0.66}
\newcommand{\beq}{\begin{equation}}
\newcommand{\eeq}{\end{equation}}

 \newtheorem{theorem}{Theorem}[section]

  \theoremstyle{definition}

\begin{document}
\baselineskip=15.5pt
\pagestyle{plain}
\setcounter{page}{1}

\begin{center}
{\LARGE \bf Renormalization Group Flow as Optimal Transport}
\vspace{0.5cm}

\textbf{Jordan Cotler$^{1,2,3}$ and Semon Rezchikov$^{4}$}

\vspace{0.4cm}

$^{1}${\it Harvard Society of Fellows, Cambridge, MA 02138 USA \\}
$^{2}${\it Black Hole Initiative, Harvard University, Cambridge, MA 02138 USA \\}
$^{3}${\it Center for Fundamental Laws of Nature, Harvard University, Cambridge, MA 02138 USA \\}
$^{4}${\it Department of Mathematics, Harvard University, Cambridge, MA 02138 USA \\}

\vspace{0.1cm}

\vspace{.1cm}

\end{center}

\begin{center}
{\bf Abstract}
\end{center}
We establish that Polchinski's equation for exact renormalization group flow is equivalent to the optimal transport gradient flow of a field-theoretic relative entropy.  This provides a compelling information-theoretic formulation of the exact renormalization group, expressed in the language of optimal transport.  A striking consequence is that a regularization of the relative entropy is in fact an RG monotone.  We compute this monotone in several examples.  Our results apply more broadly to other exact renormalization group flow equations, including widely used specializations of Wegner-Morris flow.  Moreover, our optimal transport framework for RG allows us to reformulate RG flow as a variational problem.  This enables new numerical techniques and establishes a systematic connection between neural network methods and RG flows of conventional field theories.

\newpage

\tableofcontents

\newpage

\section{Introduction}

The renormalization group is one of the central ideas in quantum field theory and statistical field theory, enabling us to understand how the effective description of a physical system changes as we tune the precision of our measurement apparatus.
There are many ways of mathematically formulating the renormalization group (RG), although a particularly illuminating way is via so-called exact renormalization group (ERG) equations, pioneered by Wilson~\cite{wilson1974renormalization} and refined by Polchinski~\cite{polchinski1984renormalization} and many others~\cite{Bagnuls:2000ae, Berges:2000ew, rosten2012fundamentals}.  ERG equations are intrinsically non-perturbative and have been used extensively in analytical and numerical investigations of RG flow over the past 40 years\mbox{~\cite{Bagnuls:2000ae, Berges:2000ew, rosten2012fundamentals}.}

A widely used ERG equation is Polchinski's~\cite{polchinski1984renormalization}, which is a functional differential equation for RG flow in a natural RG scheme.  We show that Polchinski's equation can be recast as a gradient flow of a relative entropy.  The gradient here is with respect to a functional generalization of the optimal transport metric (specifically, a version of the Wasserstein-2 metric).  The theory of optimal transport~\cite{villani2009optimal} is presently less-known to physicists, but it is a rich subject which has had a profound impact on partial differential equations and probability theory in mathematics, and optimization as well as machine learning in computer science.  We provide a review of the subject for physicists. Our results show that optimal transport is deeply ingrained in the theory of RG, enabling us to bring powerful tools from optimal transport to bear on non-perturbatively analyzing RG flows.  For instance, we precisely explain the manner in which RG flows generate entropy and clarify how this interplays with scheme-dependence; we discover a new (non-perturbative) RG monotone; and we develop a novel variational formula for RG flow which can be applied in the design of numerical methods for the renormalization group.  Our methods work for a more general class of ERG equations beyond Polchinski's, and moreover our framework provides an elegant explanation of otherwise unintuitive features of popular ERG schemes~\cite{Arnone:2003pa, Morris:2005tv, Morris:2006in}.

Let us provide a brief sketch of our results in slightly more detail.  To illustrate the basic setup of ERG equations, we consider a Euclidean scalar field theory on $\mathbb{R}^d$.  This means that we have a probability functional $P[\phi(x)] \propto e^{-S[\phi]}$ where $\phi : \mathbb{R}^d \to \mathbb{R}$ and $S[\phi]$ is the Euclidean action.  Suppose that our measurement apparatus can only probe the system down to some small distance scale $\ell$, corresponding to a UV cutoff $\Lambda \sim 1/\ell$ on the largest momenta we can access.  Now let $P_\Lambda[\phi] \propto e^{-S_\Lambda[\phi]}$ denote the probability functional corresponding to an effective description of our system given that we can only probe momentum scales less than $\Lambda$.  We are interested in how this effective description changes as we tune the value of $\Lambda$, i.e.~change the precision of our measurement apparatus.  An ERG will address this in the form of a functional differential equation
\begin{equation}
\label{E:ERGschematic1}
- \Lambda \frac{d}{d\Lambda} \, P_\Lambda[\phi] = \mathcal{F}\!\left[P_\Lambda[\phi], \frac{\delta P_\Lambda[\phi]}{\delta \phi}, \frac{\delta^2 P_\Lambda[\phi]}{\delta \phi \delta \phi},... \right]\,.
\end{equation}
The minus sign on the left-hand side indicates that we are \textit{coarse-graining} $P_\Lambda[\phi]$ in momentum space (which is done on a log scale on account of the $\Lambda \frac{d}{d\Lambda}$).  Also, the precise form of the function $\mathcal{F}$ on the right-hand side is contingent on the details of our RG scheme, or equivalently the manner in which we coarse-grain our description of the physical system in order to provide an effective description commensurate with the capabilities of our measurement apparatus.  Later on, we will precisely specify $\mathcal{F}$ for common RG schemes.

One of our main results is that Polchinski's equation can be written as
\begin{equation}
\label{E:PolchinskiS1}
- \Lambda \frac{d}{d\Lambda} \, P_\Lambda[\phi] = - \nabla_{\mathcal{W}_2} S( P_\Lambda[\phi] \, \| \, Q_{\Lambda}[\phi])
\end{equation}
where $\nabla_{\mathcal{W}_2}$ is a gradient with respect to a functional generalization of the Wasserstein-2 metric, $S(P\,\|\,Q) := \int [d\phi] P[\phi] \log(P[\phi]/Q[\phi])$ is a functional version of the relative entropy, and $Q_{\Lambda}[\phi]$ is a background probability functional which essentially defines our RG scheme.  We emphasize that our formula has the flexibility of capturing an enormous class of RG schemes.  The ingredients of our formula require further explanation, which we will provide in detail later.  Intuitively,~\eqref{E:PolchinskiS1} tells us that the coarse-graining of our theory is generated by a decrease in a relative entropy.  We will later see that the relative entropy is in fact an RG monotone; although this may seem clear from the form of~\eqref{E:PolchinskiS1}, a more detailed analysis is required which involves unpacking the definition of the gradient.

The remainder of the paper is organized as follows.  In Section~\ref{Sec:Review} we review ERG with an emphasis on Polchinksi's equation, as well as the theory of optimal transport.  In Section~\ref{Sec:RGflowOptimal} we establish equation~\eqref{E:PolchinskiS1} and a generalization pertaining to a broader class of ERG equations.  In Section~\ref{Sec:Monotones} we prove that the relative entropy appearing in our flow equations is in fact a non-perturbative RG monotone.  In Section~\ref{Sec:Examples} we compute some explicit examples of the RG monotone for both a free and interacting scalar field.  In Section~\ref{Sec:Formulations} we leverage dual formulations of optimal transport to develop a variational formula for RG flows, and then explain how it can be leveraged for new numerical methods.
Finally in Section~\ref{Sec:Discussion} we conclude with a discussion.

\section{Review of exact RG and optimal transport}
\label{Sec:Review}

Here we review pertinent tools and results about the exact renormalization group, as well as optimal transport theory.

\subsection{Exact RG}

The exact renormalization group (ERG) is a non-perturbative framework for implementing the renormalization group in quantum and statistical field theory~\cite{rosten2012fundamentals}.  In standard treatments of field theory, RG is usually implemented perturbatively via an expansion in small couplings.  By contrast, ERG provides a means to perform RG for all couplings including large couplings; in practice this is often implemented by numerical approximation schemes, but sometimes analytic methods are possible.  We begin by reviewing one of the simplest ERG equations due to Polchinski~\cite{polchinski1984renormalization} which will be our jumping off point for generalizations.

\subsubsection{Polchinki's equation}

In the spirit of Polchinski's analysis, we restrict ourselves to scalar field theory for simplicity.  We note that Polchinski's equation can be generalized to fermionic theories~\cite{Berges:2000ew, Pawlowski:2005xe, Salmhofer:2001tr} and gauge theories~\cite{Reuter:1993kw, Pawlowski:2005xe, Gies:2006wv}.

\begin{figure}
    \centering
    \includegraphics[scale = .5]{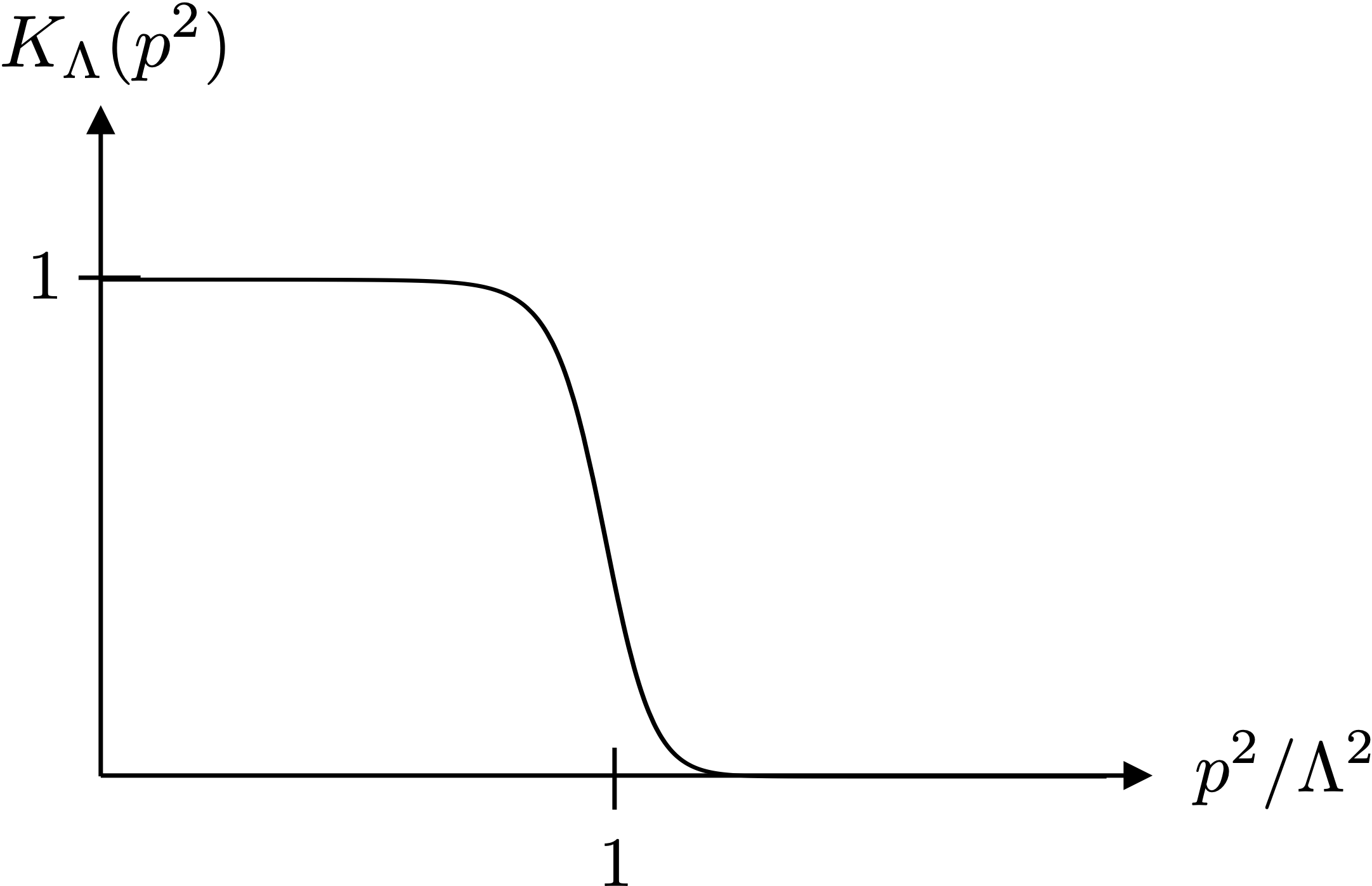}
    \caption{Depiction of a smooth cutoff function $K_\Lambda(p^2)$.}
    \label{Fig:soft1}
\end{figure}

Let us recapitulate a version of Polchinski's derivation from~\cite{polchinski1984renormalization}.  Consider a Euclidean scalar field theory with a source $J$.  We will set $\hbar = 1$ throughout.  The partition function is
\begin{equation}
\label{E:partitionJ1}
Z_\Lambda[J] := \int [d\phi] \, e^{- \frac{1}{2} \int \frac{d^d p}{(2\pi)^d} \left(\phi(p) \phi(-p) (p^2 + m^2) K_\Lambda^{-1}(p^2) + J(p) \phi(-p)\right) - S_{\text{int},\Lambda}[\phi] }\,,
\end{equation}
where $S_{\text{int},\Lambda}[\phi]$ includes interaction terms (possibly including quadratic terms which contribute to the explicit kinetic term) and where $K_\Lambda(p^2)$ is a soft cutoff function, i.e.~it is $1$ for $p^2 \lesssim \Lambda^2$ and $\approx 0$ for $p^2 \gtrsim \Lambda^2$, and $K_\Lambda^{-1}(p^2)$ denotes $1/K_\Lambda(p^2)$. This soft cutoff function ensures that correlation functions are regulated at high momentum.  For our purposes, it will be convenient for $K_\Lambda(p^2)$ to never equal zero, even if it is extremely close to zero; this way $K_\Lambda^{-1}(p^2)$ is never strictly infinite.
An example of a soft cutoff function is shown in Figure~\ref{Fig:soft1}.
Also note that the mass $m$ appearing above in~\eqref{E:partitionJ1} is the bare mass, and the couplings implicit in $S_{\text{int}, \Lambda}[\phi]$ are bare couplings.

We desire to consider some smaller scale $\Lambda_R < \Lambda$, and integrate out all modes down to $\Lambda_R$.  As such, we are only interested in computing correlation functions below the scale $\Lambda_R$, and so let us assume that our source satisfies $J(p) = 0$ for $p^2 > \widetilde{\Lambda}_R^2 - \varepsilon$ for some small $\varepsilon > 0$.  It is convenient to restrict $|m^2| \ll \Lambda_R$, i.e.~we are not integrating out the mass scale.

\begin{figure}
    \centering
    \includegraphics[scale = .5]{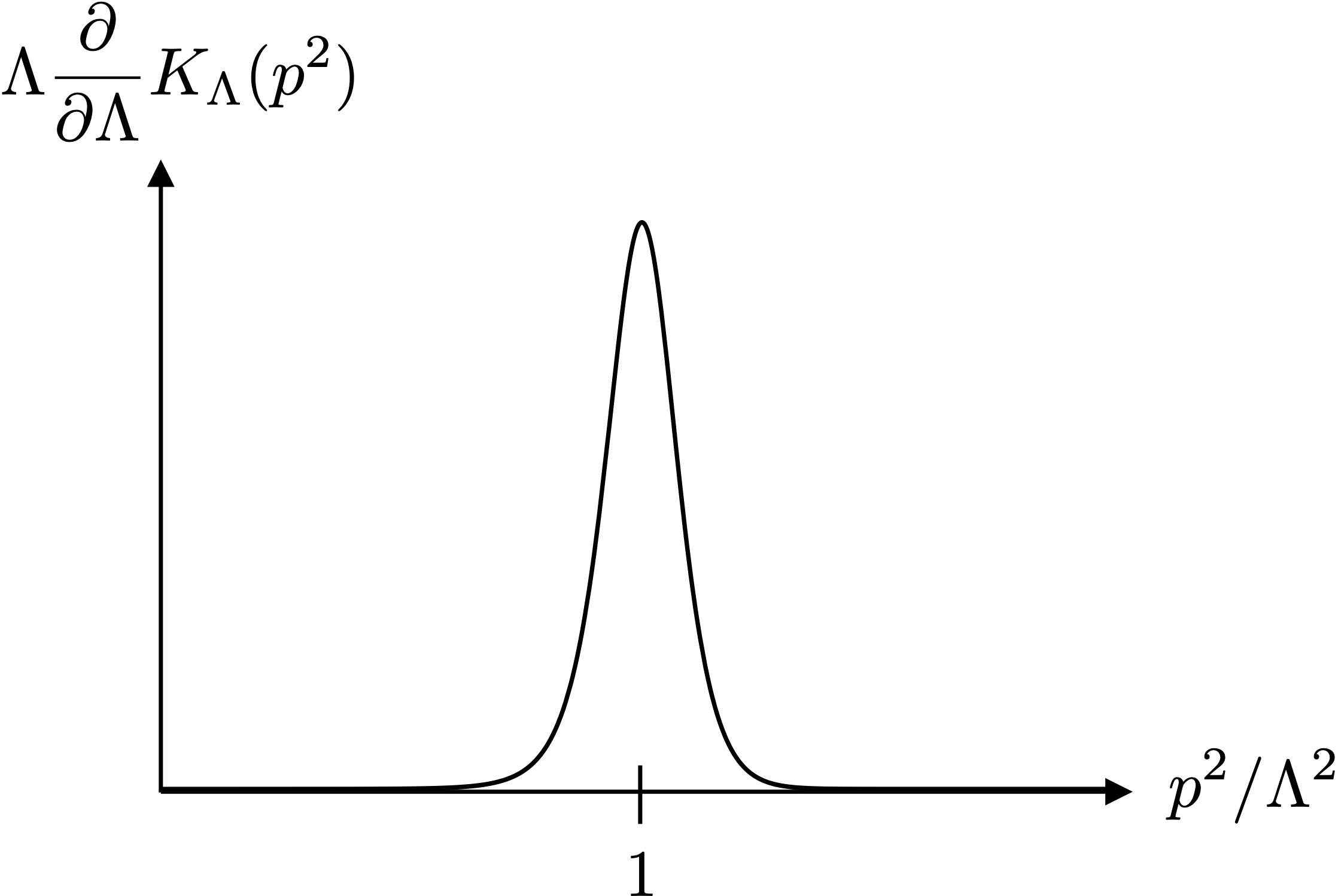}
    \caption{The derivative $\Lambda \frac{\partial}{\partial \Lambda} K_\Lambda(p^2)$ of the smooth cutoff function.}
    \label{Fig:soft2}
\end{figure}

Suppose that $\Lambda_R$ is infinitesimally smaller than $\Lambda$.  Then we would like for
\begin{equation}
\label{E:Zderivproportional}
-\Lambda \frac{d}{d \Lambda} Z_\Lambda[J] = C_\Lambda\,Z_\Lambda[J]
\end{equation}
for some constant $C_\Lambda$ only depending on $\Lambda$.  This would mean that as we change the cutoff scale $\Lambda$, which both affects the kinetic term in the action in an explicit way and the interaction terms in a way to be determined, any correlation functions below the changed scale (i.e.~generated by taking functional $J$ derivatives) stay the same.  Expanding out the left-hand side we find
\begin{align}
\label{E:Zderivativeexpanded1}
- \Lambda \frac{d}{d \Lambda} Z_\Lambda[J] &=\int [d\phi] \left( \frac{1}{2} \int \frac{d^d p}{(2\pi)^d} \,\phi(p) \phi(-p) (p^2 + m^2) \,\Lambda \frac{\partial K_\Lambda^{-1}(p^2)}{\partial \Lambda}  + \Lambda \frac{\partial S_{\text{int}, \Lambda}[\phi]}{\partial \Lambda}\right) \, e^{-S_{\Lambda}[\phi, J]}\,.
\end{align}
If we want~\eqref{E:Zderivproportional} to hold, then $\Lambda \frac{\partial S_{\text{int}}[\phi,\Lambda]}{\partial \Lambda}$ must have an appropriate form to facilitate this.  Remarkably, Polchinski found such a sufficient form which corresponds to a spatially local coarse-graining of $S_{\text{int},\Lambda}[\phi]$ upon Fourier-transforming to position space.  In particular, we will demand that $S_{\text{int},\Lambda}[\phi]$ changes with respect to $\Lambda$ via
\begin{equation}
\label{E:preRG1}
- \Lambda \frac{\partial S_{\text{int},\Lambda}[\phi]}{\partial \Lambda} = \frac{1}{2} \int d^d p \, (2\pi)^d (p^2 + m^2)^{-1}\, \Lambda \frac{\partial K_\Lambda(p^2)}{\partial \Lambda} \left\{\frac{\delta^2 S_{\text{int},\Lambda}}{\delta \phi(p) \delta \phi (-p)} - \frac{\delta S_{\text{int},\Lambda}}{\delta \phi(p)} \frac{\delta S_{\text{int},\Lambda}}{\delta \phi(-p)} \right\}\,.
\end{equation}
This is what is known as Polchinski's equation, and it is sometimes written as
\begin{equation}
\label{E:RG1}
- \Lambda \frac{\partial}{\partial \Lambda} \, e^{- S_{\text{int}, \Lambda}[\phi]} = \frac{1}{2} \int d^d p \, (2\pi)^d (p^2 + m^2)^{-1}\,\Lambda \frac{\partial K_\Lambda(p^2)}{\partial \Lambda} \,\frac{\delta^2}{\delta \phi(p) \delta \phi(-p)} \, e^{- S_{\text{int},\Lambda}[\phi]}
\end{equation}
in order to resemble a functional version of the heat equation.  Note the appearance of $\Lambda \frac{\partial K_\Lambda(p^2)}{\partial \Lambda}$ in both~\eqref{E:preRG1} and~\eqref{E:RG1}; this is localized in momentum space around $p^2 = \Lambda^2$, corresponding to a smearing kernel with scale $\sim 1/\Lambda$ in position space.  See Figure~\ref{Fig:soft2} for a depiction in momentum space.  Plugging~\eqref{E:preRG1} into~\eqref{E:Zderivativeexpanded1} and simplifying, we find
\begin{align}
\label{E:Zderivativeexpanded2}
- \Lambda \frac{d}{d \Lambda} Z_\Lambda[J] &= \left(-\frac{1}{2}\int d^d p \, \Lambda\frac{\partial  \log K_\Lambda(p^2)}{\partial \Lambda} \, \delta^d(0) \right)\, Z_\Lambda[J]
\end{align}
which has the form of the desired transformation from~\eqref{E:Zderivproportional}.

While Polchinski's equation~\eqref{E:preRG1} is formulated in terms of a functional equation for $S_{\text{int},\Lambda}[\phi]$, it will be convenient for us to recast it in terms of a functional equation for the probability functional $P_\Lambda[\phi] = e^{- S_\Lambda[\phi]}/Z_\Lambda$.  Reprocessing the above derivation we arrive at
\begin{align}
\label{E:convectiondiffusion1}
- \Lambda \frac{d}{d \Lambda} \,P_\Lambda[\phi] &=  \frac{1}{2} \int d^d p \, (2\pi)^d (p^2 + m^2)^{-1} \,\Lambda\frac{\partial K_\Lambda(p^2)}{\partial \Lambda} \,\frac{\delta^2}{\delta \phi(p) \delta \phi(-p)} \, P_\Lambda[\phi] \nonumber \\
&\qquad + \int d^d p \, \Lambda \frac{\partial \log K_\Lambda(p^2)}{\partial\Lambda} \, \frac{\delta}{\delta \phi(p)}\left(\phi(p)\,P_\Lambda[\phi]\right)
\end{align}
which has the form of a functional convection-diffusion equation.  To see the connection more clearly, we rewrite the above as
\begin{align}
\label{E:convectiondiffusion2}
- \Lambda \frac{d}{d \Lambda} \,P_\Lambda[\phi] &=  \frac{1}{2} \int d^d p \, (2\pi)^d (p^2 + m^2)^{-1} \,\Lambda\frac{\partial K_\Lambda(p^2)}{\partial \Lambda} \,\frac{\delta^2}{\delta \phi(p) \delta \phi(-p)} \, P_\Lambda[\phi] \nonumber \\
&\qquad + \frac{1}{2}\int d^d p \, (2\pi)^d (p^2 + m^2)^{-1}\, \Lambda \frac{\partial K_\Lambda(p^2)}{\partial\Lambda} \, \frac{\delta}{\delta \phi(p)}\left(\frac{2(p^2 + m^2)}{(2\pi)^d K_\Lambda(p^2)}\,\phi(p)\,P_\Lambda[\phi]\right)
\end{align}
which formally takes the same form as the finite-dimensional convection-diffusion equation
\begin{equation}
\label{E:convectiondiffusion-general}
\frac{d}{dt} \, p_t(x) = \partial_i \partial^i p_t(x) + \partial_i (v^i(x) \, p_t(x))
\end{equation}
where we identify $- \log \Lambda$ with $t$.  An example of a solution to~\eqref{E:convectiondiffusion1} (or equivalently~\eqref{E:convectiondiffusion2}) is the free theory itself; that is, $P_\Lambda[\phi] = \frac{1}{Z_{\Lambda}}\,\exp\left(-\frac{1}{2}\int \frac{d^d p}{(2\pi)^d} \, \phi(p)\phi(-p)(p^2 + m^2) K_\Lambda^{-1}(p^2)\right)$ solves Polchinski's equation.


Let us summarize the logic of Polchinski's derivation.  We explicitly differentiated $Z_{\Lambda}[J]$ by $\Lambda \frac{d}{d\Lambda}$, and then found a choice of $\Lambda \frac{\partial S_{\text{int},\Lambda}[\phi]}{\partial \Lambda}$ such that $- \Lambda \frac{d}{d\Lambda} Z_\Lambda[J] = C_{\Lambda}\, Z_\Lambda$ is satisfied for a constant $C_\Lambda$.  The suitable choice of $\Lambda \frac{\partial S_{\text{int}}[\phi]}{\partial \Lambda}$, given by the functional differential equation in~\eqref{E:preRG1}, corresponds to changing $S_{\text{int},\Lambda}$ in a manner which is localized in momentum space at scale $\Lambda$, and hence local in position space at scale $\sim 1/\Lambda$.  While Polchinski's inspired ansatz~\eqref{E:preRG1} does the job, there are in fact an infinitude of other choices which have similar properties and also render $- \Lambda \frac{d}{d\Lambda} Z_\Lambda[J] = C_{\Lambda}\, Z_\Lambda$.  These other choices correspond to alternative RG schemes than the one proposed by Polchinski.  We explore a large family of them via our discussion of the Wegner-Morris flow equation below.

\subsubsection{Wegner-Morris flow equation}

Polchinski's equation is a special case of the Wegner-Morris flow equation~\cite{wegner1974some, Morris:1999px, latorre2000exact, Morris:2005tv}.  The latter provides insights into the structure of RG flows which are obscured by Polchinski's formulation.  The Wegner-Morris equation is\footnote{We are in fact writing down a slight modification of the usual Wegner-Morris equation; the original equation only implies $- \Lambda \frac{d}{d\Lambda} Z_\Lambda = 0$, whereas we have modified the equation to allow for $- \Lambda \frac{d}{d\Lambda} Z_\Lambda = C_\Lambda \, Z_\Lambda$.}
\begin{equation}
\label{E:Wegnerflow1}
- \Lambda \frac{d}{d \Lambda} \, P_\Lambda[\phi] = \int d^d x \, \frac{\delta}{\delta \phi(x)} \left(\Psi_\Lambda[\phi, x] \,P_\Lambda[\phi]\right)
\end{equation}
and implements ERG for a scheme determined by $\Psi_\Lambda[\phi, x]$.  We note that $\Psi_{\Lambda}[\phi,x]$ will depend on $P_\Lambda[\phi]$ in a non-trivial way, which we explain below.  At first glance~\eqref{E:Wegnerflow1} does not appear to readily connect to RG flow, but its meaning will be clear shortly.

To gain some intuition for~\eqref{E:Wegnerflow1}, it is useful to compare with a finite-dimensional analog.  This would be the equation for $p_t$ given by
\begin{equation}
\frac{d}{dt}\,p_t(x) + \partial_i (V^i(p_t, x) \, p_t) = 0\,.
\end{equation}
In this equation the vector field $V^i$, the analog of $\Psi_\Lambda$ in the Wegner-Morris flow, is chosen to depend not just on the coordinate position $x$ but also on the entire probability distribution $p_t$.  It is natural for $V^i$ to satisfy $V^i(p_t,x) = \partial^i W(p_t, x)$, namely for $V^i$ to have a potential $W$.  This gives us
\begin{equation}
\label{E:finitedimW}
\frac{d}{dt}\,p_t(x) + \partial_i (\partial^i W(p_t, x) \, p_t) = 0\,.
\end{equation}
We will find an analogue of this potential in the Wegner-Morris flow equation for many cases of interest.

Equation~\eqref{E:Wegnerflow1} has several features which illuminate its meaning.  First, performing the functional integral of both sides of~\eqref{E:Wegnerflow1} with respect to $\phi(x)$ and noting that $\Psi_\Lambda[\phi, x] \,P_\Lambda[\phi]$ goes to zero for large $\phi(x)$, we immediately see that $- \Lambda \frac{d}{d\Lambda} \int [d\phi] \, P_\Lambda[\phi] = 0$ and so the flow equation preserves probability.  More generally, the meaning of~\eqref{E:Wegnerflow1} is that as the scale $\Lambda$ changes the flow induces the field reparameterization
\begin{equation}
\label{E:field-reparametrization}
\phi'(x) = \phi(x) + \frac{\delta \Lambda}{\Lambda}\, \Psi_\Lambda[\phi, x]\,.
\end{equation}
This means that the probability functional is simply reparameterized by the flow, and so probability is clearly conserved and positivity of the probability density is maintained.  As explained in~\cite{latorre2000exact}, essentially all RG schemes (with a soft cutoff) can be cast into the form of the Wegner-Morris flow equation above.  In all schemes $\Psi_\Lambda$ instantiates field redefinitions which are localized in momentum space near scale $\Lambda$, i.e.~we are reparameterizing the field at or near the cutoff scale.  We will henceforth refer to $\Psi_\Lambda$ as the reparameterization kernel.

A common form of $\Psi_\Lambda[\phi, x]$ is given by~\cite{Arnone:2002cs, Arnone:2003pa, Arnone:2005fb, Morris:2005tv, Morris:2006in, rosten2012fundamentals}
\begin{equation}
\label{E:Psieq1}
\Psi_\Lambda[\phi, x] = -\int d^d y \, \frac{1}{2}\,\dot{C}_\Lambda(x-y)\, \frac{\delta \Sigma_\Lambda[\phi]}{\delta \phi(y)}\,,
\end{equation}
where $\dot{C}_\Lambda(x-y)$ is called the ERG kernel\footnote{We have chosen a different sign convention than the usual literature, namely $\dot{C}_{\Lambda,\,\text{us}} = - \dot{C}_{\Lambda,\,\text{them}}$.  This extra minus sign will make some of our later formulas more intuitive.} which satisfies $\dot{C}_\Lambda(x-y) \geq 0$, and
\begin{equation}
\label{E:Sigma1}
\Sigma_\Lambda[\phi] := S_\Lambda[\phi] - 2 \hat{S}_\Lambda[\phi]
\end{equation}
where $S_\Lambda[\phi]$ is the action appearing in $P_\Lambda[\phi] = e^{-S_\Lambda[\phi]}/Z_\Lambda$ and $\hat{S}_\Lambda[\phi]$ is another action called the `seed action'.  The multiplicative factor of $2$ in front of the seed action is conventional.  In its present form, the meaning of the seed action is physically obscure.    Fortunately, our optimal transport analysis later on will elucidate its meaning.  Notice that $\Psi_\Lambda[\phi, x]$ is a gradient of $\Sigma_\Lambda[\phi]$, where $\frac{1}{2}\,\dot{C}_\Lambda(x-y)$ plays the role of an inverse metric, and so in this setting the Wegner-Morris flow equation~\eqref{E:Wegnerflow1} takes the form of the finite-dimensional equation~\eqref{E:finitedimW}.

Importantly, we can reproduce the Polchinski's equation with the choices
\begin{align}
\dot{C}_\Lambda(p^2) &= (2\pi)^d (p^2 + m^2)^{-1} \,\Lambda \frac{\partial K_\Lambda(p^2)}{\partial \Lambda} \\
\hat{S}_\Lambda &= \frac{1}{2}\int \frac{d^d p}{(2\pi)^d}\,(p^2 + m^2)\, K_\Lambda^{-1}(p^2)\,\phi(p) \phi(-p)\,,
\end{align}
here expressed in momentum space.\footnote{In the equation for $\dot{C}_\Lambda(p^2)$, the right-hand side is greater than or equal to zero.  Since $\dot{C}_\Lambda(p^2)$ is continuous, Bochner's theorem implies that its Fourier transform $\dot{C}_\Lambda(x-y)$ is likewise greater than or equal to zero.}  Notice that $\hat{S}$ is just an action for a free massive scalar field with the same initial bare mass as our scalar field theory of interest.

An initially puzzling feature of Wegner-Morris flow is that~\eqref{E:field-reparametrization} can be inverted if $\Psi_\Lambda[\phi, x]$ is well enough behaved.  This would mean that the exact RG flow is invertible.  However, we often think of RG as being non-invertible, perhaps the most famous example being Kadanoff's block spin decimation for spin systems (see e.g.~\cite{kadanoff1966scaling, kadanoff2000statistical}).  For continuum field theories, exact RG flows are typically invertible, although the inversion is ill-conditioned.  As an example close in spirit to Kadanoff's block spin methods, suppose that our RG flow is prescribed by the coarse-graining\footnote{This can be written in the form of the Wegner-Morris flow equation~\eqref{E:Wegnerflow1}, albeit with $\Psi_\Lambda$ taking a form different from the ansatz class in~\eqref{E:Psieq1},~\eqref{E:Sigma1}.  See~\cite{rosten2012fundamentals} for a discussion.} $P_\Lambda[\phi] = \int [d\psi] \, \delta[\phi - b_\Lambda[\psi]] \, P_{\Lambda_0}[\psi]$ where $\Lambda_0$ is the initial RG scale and $\Lambda \leq \Lambda_0$.  Here $b_\Lambda[\psi](x) := \int d^d y \, f_\Lambda(x-y) \, \psi(y)$, where $f_\Lambda(x-y)$ is a smearing kernel with width $\sim 1/\Lambda$ in position space.  Perhaps $f_\Lambda(x-y)$ is a Gaussian distribution, or a $d$-dimensional unit box function (which has compact support).  So we are performing a continuum version of Kadanoff's procedure.  However, a key difference is that the smearing $\int d^d y \, f_\Lambda(x-y) \, \psi(y)$ is invertible in the continuum as can be seen by transforming to Fourier space to get $f_\Lambda(p) \psi(p)$ and dividing by $f_\Lambda(p)$.  Indeed, if $f_\Lambda(x-y)$ is a Gaussian, then so is its Fourier transform; dividing by a Gaussian is well-defined, albeit ill-conditioned since we are dividing by very small numbers in the tail regions.  Likewise the Fourier transform of a box function is the product of sinc functions, and division by them is likewise ill-conditioned.\footnote{Here we also need to be careful about dividing by zero at isolated points, but this can be dealt with if the fields $\psi$ belong to a sufficiently nice function class. Relatedly, the invertibility of block spin renormalization fails to apply to the discretized lattice setting due the function class corresponding to latticized fields.} 

More broadly, even when we perform exact versions of the more standard Wilsonian RG, the flow is only in general invertible if we keep the infinitely many irrelevant terms in the action generated by the flow.

\subsection{Optimal Transport}

As discussed above, Polchinski's equation for $P_\Lambda$ is an infinite-dimensional convection-diffusion equation, which can be thought of as a generalized form of heat flow.  The RG monotones we present later on will be analogs of the entropy of a distribution.  The fact that the entropy of a distribution is monotone along heat flows was already known to Gibbs.  However, the understanding that the entropy functional generates heat flow under the Wasserstein metric required a synthesis~\cite{otto_dissipative_evolution} of ideas about \emph{optimal transport}.  This synthesis occurred relatively recently in the 90's, in the work of Otto, Benamou-Brenier, and many others.  We will review some of these developments here.

At a high level, the problem of optimal transportation is to determine an optimal method for moving and rearranging a given mass distribution into a desired mass distribution, given a cost for moving mass across a specified distance. In the next three subsections, we will review the basic mathematical formalization, discuss fundamental results about this problem, and explain how it connects with heat flow. Beyond this connection, there is a rich theory connecting optimal transport with probability theory and mathematical physics, and we will provide a short guide to relevant literature for interested readers.

\subsubsection{Monge and Kantorovich formulations}

Given a space $X$ and a pair of probability or mass distributions $p, q$ on $X$, the \emph{Monge formulation} of the optimal transport problem asks to find a (measurable) \emph{transport function} $T: X \to X$ such that:
\begin{figure}
    \centering
    \includegraphics[scale = .4]{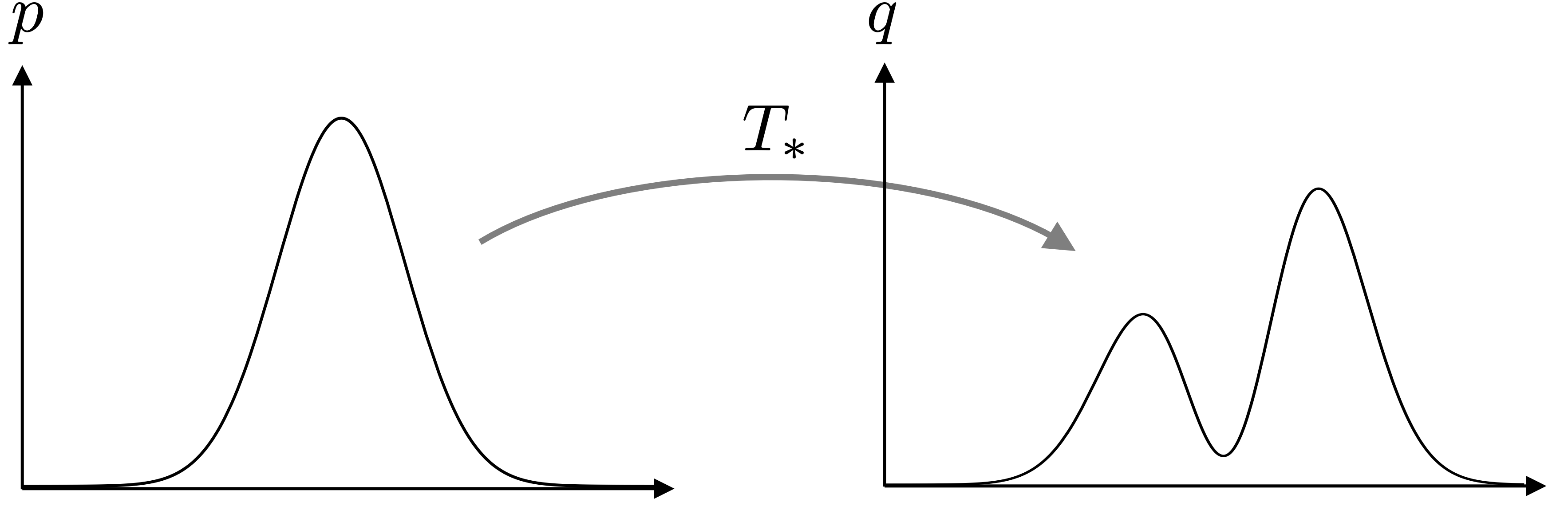}
    \caption{Schematic of the probability mass distributions $p$ and $q$, and the map $T_*$ between them.
    \label{Fig:optimal1}}
\end{figure}
\begin{enumerate}
    \item The pushforward of $p$ under $T$ is $q$, i.e.~$T_* p = q$\,; equivalently $\int_{T^{-1}(S)} dx \, p(x) = \int_{S} dx\, q(x)$ for every measurable set $S$; and 
    \item The transport function minimizes the total cost 
    \begin{equation}
    \label{E:monge-functional}
        \textsf{M}[T] = \int_X dx\, p(x)\, c(x, T(x))
    \end{equation}
    for some cost function $c: X \times X \to \R$. 
\end{enumerate}
A natural choice for the cost function is $c(x,y) = d(x,y)^2$, where $d$ is a distance function on $X$.  A depiction of the mapping $T_* p = q$ can be seen in Figure~\ref{Fig:optimal1}.

The constraint $T_*p = q$ is highly nonlinear, making the existence of a solution non-obvious. For concreteness, suppose that $X=\R^n$ 
and $T$ is a smooth function.  We let $T_j$ denote the $j$th coordinate output of $T$.  Then the constraint can be written as
\begin{equation}
\label{eq:monge-constraint}
    q(T(x))\, |\det(\partial_i T_j(x))| = p(x)\,.
\end{equation}
This nonlinear constraint above makes it difficult to establish the existence of solutions to the Monge problem via methods from the calculus of variations.  Worse, solutions to the Monge problem no longer exist once the distributions are not smooth: if the distributions $p, q$ are sums of delta functions, i.e.~$p(x) = \sum_i p_i \,\delta(x - a_i)$ while $q(x) = \sum_i q_i \,\delta(x - b_i)$, then for generic choices of supports $\{a_i\}, \{b_j\}$, it is clear that no transport map $T$ exists.  For instance, if $p$ is supported on one point and $q$ is supported on two points, there is no transport map $T$ such that $T_* p = q$.

To better understand the Monge problem, it is convenient to first solve a relaxation known as the Kantorovich problem. In the Kantorovich problem, one searches for a positive measure $\pi$ on $X \times X$ such that:
\begin{enumerate}
    \item The pushforward of $\pi$ to $X$ is $p$, and the pushforward of $\pi$ to $Y$ is $q$ (i.e.~$\int_X dy\, \pi(x,y) = p(x)$ and $\int_X dx\, \pi(x,y) = q(y)$); and 
    \item The measure $\pi$ minimizes
    \begin{equation}
        \mathsf{K}(\pi) = \int_{X \times X} \!dx\,dy\, \pi(x,y) \,c(x,y)\,. 
    \end{equation}
\end{enumerate}
The interpretation of $dx\,dy\,\pi(x,y)$ is that it is the infinitesimal amount of mass at $x$ which is transported to $y$.  If we set $\pi_{x,y} = p(x) \delta(y - T(x))$ then it is clear that $\mathsf{K}(\pi) = \mathsf{M}(T)$.  Thus, candidate solutions to the Monge problem give candidate solutions to the Kantorovich problem. However, the Kantorovich problem is much easier, as it is a problem in infinite dimensional \textit{convex} optimization. Indeed, the function $\textsf{K}(\pi)$ is a linear function on the convex cone of positive measures on $X \times X$ and the constraints arising from $p$ and $q$ are also linear. Discretizing this optimization problem yields a familiar finite-dimensional linear program: if $p(x) = \sum_i p_i \, \delta(x - a_i)$, $q(y) = \sum_j q_j \delta(y - b_j)$, and $\pi(x,y) = \sum_{i, j} \pi_{ij} \delta(x-a_i) \delta(y-b_j)$, then the Kantorovich problem immediately reduces to 
\begin{equation}
\text{Minimize}\quad \sum_{i, j} \pi_{ij} \,c(a_i, b_j) \quad \text{subject to} \quad \pi_{ij} \geq 0\,,\quad \sum_j \pi_{ij} = p_i\,, \quad \sum_i \pi_{ij} = q_j\,.
\end{equation}

Despite that fact that the Kantorovich problem is a relaxation of the Monge problem, in a large class of cases solutions to the Kantorovich problem actually arise from solutions to the Monge problem:
\begin{theorem}
If $p(x)$ and $q(x)$ are smooth functions having support on all of $\R^n$, then the Monge problem with $c(x,y) = |x-y|^2$ has a smooth solution; indeed, we have 
\begin{equation}
\label{eq:Brenier-potential}
    T_i(x) = \partial_i f(x)
\end{equation}
for some smooth convex function $f: \R^n \to \R$.
\end{theorem}
\noindent This result requires the development of a significant amount of mathematics: it follows from a combination of duality for the Kantorovich problem, Brenier's theorem~\cite{brenier_polar_factorization}, and Cafarelli's regularity theory~\cite{Caffarelli1992} for solutions to the Monge-Amp\`{e}re equation. To explain the proof of this theorem would take us too far afield, although~\cite[Chapters~2-4]{villani2009optimal} gives a good introduction. We only note that once the existence of a function satisfying (\ref{eq:Brenier-potential}) is established by the duality theory, one concludes by the constraint (\ref{eq:monge-constraint}) that $f$ satisfies the equation 
\begin{equation}
    \det\!\left(\text{Hess}\,f(x)\right) = \frac{p(x)}{q(\nabla f(x))}
\end{equation}
which is the form of the Monge-Amp\`{e}re equation that appears in this setting.

\subsubsection{Wasserstein distance}

Since we will be primarily interested in cases for which the space $X$ is a metric space (and often a Riemannian metric space), we will henceforth denote the space by $M$.  For the quadratic cost $c(x,y)=|x-y|^2$ where $x,y$ are Cartesian coordinates on Euclidean space $M = \R^n$,
the optimum value of $\textsf{K}(\pi)$ in the Kantorovich problem is called the \textit{Wasserstein-2 distance} $\mathcal{W}_2(p_1, p_2)$.  (This is alternatively called the $L^2$-Wasserstein distance.)  
The distance can be written as
\begin{equation}
\label{E:globalW2}
\mathcal{W}_2(p_1,p_2) := \left(\inf_{\pi \in \Gamma(p_1, p_2)} \int_{M \times M} dx \, dy \,\pi(x,y) \, |x-y|^2 \right)^{1/2}
\end{equation}
where $\Gamma(p_1, p_2)$ is the space of probability distributions $\pi(x,y)$ on $M \times M$ such that $\int_M dy \, \pi(x,y) = p(x)$ and $\int_M dx \, \pi(x,y) = q(y)$.
This metric distance on the space of probability distributions, and in particular various path integral generalizations of it, will play a central role in our analyses.

\subsubsection{Otto calculus}
\label{Subsec:Otto1}

To explain the connection between heat flow and optimal tranport, we first recall how to view heat flow as a gradient flow with respect to the usual $L^2$ metric. 

\paragraph{Heat flow as gradient flow of Dirichlet energy.}
For a function $F: \M \to \R$ on a Riemannian manifold $\mathcal{M}$ with metric $\langle \cdot, \cdot \rangle$, the gradient of $F$ at $x_0 \in \M$ is the vector $\Grad F(x_0)$ such that 
\begin{equation}
\label{eq:gradient-definition}
   \frac{d}{dt} F(x(t))\Big|_{t = 0} = \Big\langle \Grad F(x_0), \frac{\partial}{\partial t} x(t)\Big\rangle \Big|_{t = 0}
\end{equation}
for every curve $x(t) \in \mathcal{M}$ with $x(0) = x_0$.

For our purposes, we let $\mathcal{M} = \text{dens}(M)$ be the space probability densities on a manifold $M$, where we suppose $M$ is equipped with a volume form $dV$.  That is, $\text{dens}(M)$ is an infinite-dimensional manifold defined by
\begin{equation}
    \text{dens}(M) := \left\{ p \in C^\infty(M) \,\Bigg| \,\,p \geq 0\,,\,\int dV\,p = 1 \right\}\,.
\end{equation}
The tangent space at $p \in \text{dens}(M)$ is 
\begin{equation}
    T_p\, \text{dens}(M) = \left\{ \eta \in C^\infty(M) \,\Bigg|\, \int dV\,\eta = 0\right\}\,.
\end{equation}
We can equip each tangent space $T_p \,\text{dens}(M)$ with a Riemannian metric 
\begin{equation}
\label{E:L2metric1}
    \langle \eta_1, \eta_2 \rangle_{L^2} = \int dV\,\eta_1 \eta_2\,.
\end{equation}
This corresponds to the $L^2$ inner product on functions on $M$.  Defining the Dirichlet energy functional as
\begin{equation}
\mathcal{E}[p] := \frac{1}{2}\int dV\,|\Grad p|^2\,,
\end{equation}
we can compute its gradient with respect to the infinite-dimensional $L^2$ metric in~\eqref{E:L2metric1} using (\ref{eq:gradient-definition}).  In particular, let $\rho(t)$ be a differentiable path through $\text{dens}(M)$ such that $\rho(0) = p$.
Then
\begin{align}
    \frac{d}{dt} \,\mathcal{E}[\rho(t)] &= \int dV\, \Grad \rho \cdot \Grad \frac{\partial}{\partial t} \,\rho \nonumber \\
    &= -\int dV\,\Delta \rho \,\frac{\partial}{\partial t}\, \rho \nonumber \\
    &= \Big\langle \!- \Delta \rho,\, \frac{\partial}{\partial t}\, \rho\Big\rangle_{L^2}\,. 
\end{align}
Evaluating the above at $t=0$ and comparing with~\eqref{eq:gradient-definition}, we read off that
\begin{equation}
\Grad_{L^2} \mathcal{E}[p] = - \Delta p\,.
\end{equation}
It follows that the heat equation $\frac{\partial}{\partial t} \,p = \Delta p$ is the negative gradient flow of the Dirichlet energy functional $\mathcal{E}$, namely
\begin{equation}
\label{E:heatgradientflow1}
\frac{\partial}{\partial t}\,p(x,t) = - \Grad_{L^2} \mathcal{E}[p(x,t)]\,.
\end{equation}
This in fact implies that the Dirichlet energy monotonically decreases along the heat flow.

\paragraph{Wasserstein distance and the gradient flow of entropy.}
Another monotone for the heat flow is given by the differential entropy
\begin{equation}
\label{eq:entropy}
    S[p] := - \int dV\,p \log(p)\,. 
\end{equation}
We will have more to say about this quantity in Subsection~\ref{Subsec:commentsandinterp1}.  By analogy with~\eqref{E:heatgradientflow1} above, we might ask if there is any Riemannian metric $g$ on $\text{dens}(M)$ such that the heat equation can be written as $\frac{\partial}{\partial t}\,p = \nabla_g S[p]$\,?  In other words, is there some (natural) metric on the space of probability distributions for which the heat equation is the gradient flow of the differential entropy?

Remarkably, the answer yes -- this was discovered by Otto~\cite{otto_dissipative_evolution} and widely exploited by subsequent researchers in partial differential equations and probability theory.  In fact, there are a large collection of entropy-like monotones $\widetilde{S}$ which have associated metrics $\widetilde{g}$ on $\text{dens}(M)$ such that the heat equation can be written as $\frac{\partial}{\partial t}\,p = \nabla_{\widetilde{g}} \,\widetilde{S}[p]$.  All of these metrics have deep connections to optimal transport.  Since we will be interested in the particular case of the differential entropy, we will not discuss these related entropic gradient flow formulations here.

We now turn to constructing the metric $g$ on $\text{dens}(M)$ such that $\frac{\partial}{\partial t}\,p = \nabla_g S[p]$.  To write the metric in the most transparent way, an isomorphism of the tangent space $T_p \,\text{dens}(M)$ is required.  Given a tangent vector $\eta \in T_p \,\text{dens}(M)$, we can solve for a $\bar{\eta}$ satisfying
\begin{equation}
\label{eq:dualizing-equation}
    \Grad \cdot ( p \Grad \bar{\eta})= \eta\,.
\end{equation}
The solution is unique up to addition of a constant, and so we get an identification $\eta \leftrightarrow \bar{\eta}$ which we notate by the isomorphism
\begin{equation}
T_p\,\text{dens}(M) \simeq \overline{T_p\,\text{dens}(M)} := \{ \bar{\eta} \in C^\infty(M)\}/\{\text{constants}\}\,. 
\end{equation}
Using this identification we define the Riemannian metric 
\begin{equation}
\label{eq:infinitesimal-wasserstein-metric}
\langle \eta_1, \eta_2 \rangle_{\mathcal{W}_2} := \int dV\,p\,\Grad \bar{\eta}_1\cdot \Grad \bar{\eta}_2 = -\int dV\, \eta_1 \bar{\eta}_2 = -\int dV\, \bar{\eta}_1 \eta_2 
\end{equation}
where the last two equalities can be checked via integration by parts.  This metric is in fact the infinitesimal form of the Wasserstein-2 distance $\mathcal{W}_2$. A rigorous argument establishing this fact is given in~\cite[Lemma 4.3]{otto2005eulerian}; we will explain the heuristic connection in Appendix~\ref{App:heuristic}.

Now let us show that $\nabla_{\mathcal{W}_2} S[p] = \Delta p$.  Let $\rho(t)$ be a path through $\text{dens}(M)$ with $\rho(0) = p$, and define $\eta := \frac{d}{dt} \rho(t)\big|_{t = 0}$ which is definitionally an element of $T_p\,\text{dens}(M)$.  Let $\bar{\eta}$ be the corresponding solution to~\eqref{eq:dualizing-equation}.  Then we compute
\begin{align}
\frac{d}{dt}\,S[\rho(t)]\Big|_{t = 0} &= -\int dV\,\eta(\log p + 1) \nonumber \\
&= -\int dV \, \nabla \cdot(p \nabla \bar{\eta})(\log p + 1) \nonumber \\
&= \int dV \, \nabla \bar{\eta} \cdot \nabla p \nonumber \\
&= - \int dV \, \bar{\eta}\cdot \Delta p \nonumber \\
&= \langle \Delta p, \, \eta \rangle_{\mathcal{W}_2} \nonumber \\
&= \Big\langle \Delta p,\, \frac{\partial}{\partial t}\,\rho\Big\rangle_{\!\mathcal{W}_2}\,\Big|_{t = 0}
\end{align}
and so comparing with~\eqref{eq:gradient-definition} we indeed find
\begin{equation}
    \nabla_{\mathcal{W}_2} S[p] = \Delta p\,.
\end{equation}
Then the heat equation can be written as
\begin{equation}
\frac{\partial}{\partial t}\,p(x,t) = \nabla_{\mathcal{W}_2} S[p]\,.
\end{equation}
Thus, the heat flow is the gradient flow of the differential entropy (\ref{eq:entropy}) with respect to the Wasserstein-2 metric.  While it was known to Gibbs that entropy is a heat flow monotone, the above equation clarifies that in fact heat flow is \textit{completely governed} by the entropy, with optimal transport playing a central role in this formulation.

\subsubsection{A guide to further literature}
In the rest of this paper, we will exploit formal, infinite-dimensional analogues of the optimal-transport formulation of heat flow to study the renormalization group.  We expect that that there are further profitable connections to be made between the rich mathematics of optimal transport and the structure of the renormalization group, and we view the present work as an initial study. 

The lecture notes~\cite{villani2009optimal} are a very readable mathematical introduction to the subject of optimal transport, and the original paper~\cite{otto_dissipative_evolution} remains  full of geometric insight. The papers~\cite{brenier_polar_factorization, benamou_brenier, Caffarelli1992} mentioned above are all fundamental. The book~\cite{villani2009optimal} covers connections to Ricci curvature, while the review article~\cite{mcann_lectures} summarizes applications in PDE and applied mathematics. The logarithmic Sobolev inequalities proven in~\cite{gross_log_sobolev} were reproven using optimal transport in~\cite{ottovillani} and had a dramatic impact on probability theory; they were originally motivated by problems in constructive quantum field theory and so it is not surprising that the ideas should come full circle. A very recent application of Polchinski's equation to constructive quantum field theory can be found in~\cite{bauerschmidt_bodineau}. We note also that ideas around the logarithmic Sobolev inequalities together with the fact that Ricci flow is renormalization group flow for a $\sigma$-model was a stated motivation for Perelman's work on the Poincar\`e conjecture~\cite{perelman2002entropy}; following this idea, McCann-Topping~\cite{mcann_ricci} began an ongoing research program founding Ricci flow in ideas based on optimal transport.

\section{RG flow as an optimal transport gradient flow}
\label{Sec:RGflowOptimal}

\subsection{Deriving the optimal transport gradient flow equation for RG}

Since Polchinski's equation~\eqref{E:convectiondiffusion1} is a special case of the Wegner-Morris flow equation~\eqref{E:Wegnerflow1}, we find it prudent to derive our optimal transport equation for the latter.  Suppose we intend to flow a Euclidean field theory with probability functional $P_\Lambda[\phi] = e^{-S_\Lambda[\phi]}/Z_{P,\Lambda}$.  Recall the Wegner-Morris flow equation
\begin{equation*}
- \Lambda \frac{d}{d \Lambda} \, P_\Lambda[\phi] = \int d^d x \, \frac{\delta}{\delta \phi(x)} \left(\Psi_\Lambda[\phi, x] \,P_\Lambda[\phi]\right)
\end{equation*}
where will adopt the functional forms in~\eqref{E:Psieq1} and~\eqref{E:Sigma1} for the reparameterization kernel $\Psi_\Lambda$, namely
\begin{equation*}
\Psi_\Lambda[\phi, x] = -\int d^d y \, \frac{1}{2}\,\dot{C}_\Lambda(x-y) \, \frac{\delta \Sigma_\Lambda[\phi]}{\delta \phi(y)}\,, \qquad \Sigma_\Lambda[\phi] = S_\Lambda[\phi] - 2 \hat{S}_\Lambda[\phi]
\end{equation*}
where $\dot{C}_{\Lambda}(x-y) \geq 0$.  Now we define a Riemannian metric on the tangent space to the space of probability functionals which we will later explain is the infinitesimal version of a functional $\mathcal{W}_2$ metric.  We let
\begin{align}
\label{E:functionalRiemannian1}
\langle \delta P_1[\phi], \, \delta P_2[\phi] \rangle_{\mathcal{W}_2} = \frac{1}{2} \int [d\phi] \, P[\phi] \int d^d x \, d^d y \, \dot{C}_\Lambda(x-y)\, \frac{\delta\Phi_1[\phi]}{\delta \phi(x)} \frac{\delta\Phi_2[\phi]}{\delta \phi(y)}
\end{align}
where we define $\Phi_i[\phi]$ for $i=1,2$ via the functional differential equations
\begin{equation}
\label{E:Phieq1}
\delta P_i[\phi] - \frac{1}{2} \int d^d x \, d^d y \, \dot{C}_\Lambda(x-y)\, \frac{\delta}{\delta \phi(x)}\left(P[\phi]\,\frac{\delta\Phi_i[\phi]}{\delta \phi(y)} \right) = 0\,.
\end{equation}
Analogous to the finite-dimensional heat flow setting, the $\Phi_i$'s are only specified by the above equation up to additive functions not depending on $\phi$.  Note that since $\dot{C}_{\Lambda}(x-y) \geq 0$, the norm induced by the metric is automatically greater than or equal to zero.
Similar to Otto's calculation we can perform an integration by parts in~\eqref{E:functionalRiemannian1} to obtain the more compact expressions
\begin{equation}
\label{E:functionalRiemannian2}
\langle \delta P_1[\phi], \, \delta P_2[\phi] \rangle_{\mathcal{W}_2} = - \int [d\phi] \,\delta P_1[\phi] \, \Phi_2[\phi] = - \int [d\phi] \, \Phi_1[\phi] \, \delta P_2[\phi]\,.
\end{equation}
Co-opting the results of Otto~\cite{otto2005eulerian} and generalizing them appropriately to our setting, we have that our metric is the infinitesimal form of the distance
\begin{align}
\label{E:functionaltotalW2}
&\mathcal{W}_2(P_1, P_2) \nonumber \\
&\,\,\, := \bigg(\inf_{\Pi \in \Gamma(P_1, P_2)}  2 \int [d\phi_1]\,[d\phi_2]\,\Pi[\phi_1, \phi_2] \int d^d x \, d^d y\, \dot{C}_\Lambda^{-1}(x,y) \left(\phi_1(x) - \phi_2(x)\right)\left(\phi_1(y) - \phi_2(y) \right)\!\bigg)^{1/2}
\end{align}
where $\Gamma(P_1, P_2)$ is the space of probability functionals $\Pi[\phi_1,\phi_2]$ such that $\int [d\phi_2] \, \Pi[\phi_1, \phi_2] = P[\phi_1]$ and $\int [d\phi_1] \, \Pi[\phi_1, \phi_2] = P_2[\phi_2]$.  Above $\dot{C}_\Lambda^{-1}(x,y)$ is the inverse of the kernel $\dot{C}_\Lambda(x,y)$ in the sense that $\int d^d z \, \dot{C}_\Lambda^{-1}(x,z) \,\dot{C}_\Lambda(z,y) = \delta^d(x-y)$; since in our setting $\dot{C}_\Lambda(x,y) = \dot{C}_\Lambda(x-y)$, in momentum space the kernel $\dot{C}_\Lambda(p^2)$ has as its inverse $\dot{C}_\Lambda^{-1}(p^2) = 1/\dot{C}_\Lambda(p^2)$.  The distance $\mathcal{W}_2(P_1, P_2)$ represents the minimum cost of `transporting' $P_1$ into $P_2$ (or vice-versa) where the cost is given by an $L^2$ penalty on rearranging field degrees of freedom away from the spatial scale $\ell \sim 1/\Lambda$.

We are now almost ready to state our main result, and then subsequently derive it.  Define the probability functional
\begin{equation}
\label{E:Q1}
Q_\Lambda[\phi] := \frac{e^{-2\hat{S}_\Lambda[\phi]}}{Z_{Q,\Lambda}}
\end{equation}
where $Z_{Q,\Lambda} = \int [d\phi]\,e^{-2\hat{S}_\Lambda[\phi]}$, and let the functional relative entropy be
\begin{equation}
\label{E:relentropy1}
S(P[\phi] \,\|\, Q[\phi]) := \int [d\phi] \, P[\phi] \log\left(\frac{P[\phi]}{Q[\phi]}\right)\,.
\end{equation}
Then we have the remarkable formula
\begin{equation}
\label{E:boxed1}
\boxed{- \Lambda \frac{d}{d\Lambda} P_\Lambda[\phi] = - \nabla_{\mathcal{W}_2} S( P_\Lambda[\phi]\,\|\,Q_\Lambda[\phi])}
\end{equation}
which is equivalent to the Wegner-Morris flow equation~\eqref{E:Wegnerflow1}.  To establish this connection, we need to show that $- \nabla_{\mathcal{W}_2} S( P_\Lambda[\phi]\,\|\,Q_\Lambda[\phi])$ equals $\int d^d x \, \frac{\delta}{\delta \phi(x)} (\Psi_\Lambda[\phi, x] \, P_\Lambda[\phi])$\,.

For any $\mathcal{F}[P[\phi]]$ which takes probability functionals to the real numbers, the differential-geometric definition of the gradient $\nabla_{\mathcal{W}_2} \mathcal{F}[P]$ is given by
\begin{equation}
\label{E:functionalalgradient1}
\langle \nabla_{\mathcal{W}_2} \mathcal{F}[P], \delta P \rangle_{\mathcal{W}_2} = \int [d\phi] \, \frac{\delta \mathcal{F}[P]}{\delta P} \, \delta P[\phi]\,.
\end{equation}
A slightly unusual feature of the right-hand side is that $\frac{\delta \mathcal{F}[P]}{\delta P}$ is not an ordinary functional derivative but rather a functional-of-a-functional derivative, i.e.~a derivative with respect to the functional $P[\phi]$.  In our case, we choose $\mathcal{F}[P_\Lambda] := S( P_\Lambda \|\,Q_\Lambda)$; then computing the right-hand side of~\eqref{E:functionalalgradient1} we obtain
\begin{align}
\int [d\phi] \left(\log P_\Lambda[\phi] + 1 - \log Q_\Lambda[\phi]\right) \delta P[\phi] &= \int [d\phi] \left(- S_{\Lambda}[\phi] - \log Z_P + 1 + 2\hat{S}_\Lambda[\phi] + \log Z_Q\right) \delta P[\phi] \nonumber \\
&= \int [d\phi] \left(- S_{\Lambda}[\phi] + 2\hat{S}_\Lambda[\phi] \right) \delta P[\phi] \nonumber \\
&= - \int [d\phi] \, \Sigma_\Lambda[\phi] \, \delta P[\phi]\,.
\end{align}
In going from the first to second line we used $\int [d\phi] \, \delta P[\phi] = 0$ since this is a property of elements of the tangent space to probability functionals so that $\int [d\phi] (P[\phi] + \delta P[\phi]) = 1$.  Next we use~\eqref{E:Phieq1} to rewrite $\delta P$ in terms of a $\Phi$ field, giving us
\begin{align}
- \frac{1}{2} \int [d\phi] \, \Sigma_\Lambda[\phi] \int d^d x \, d^d y \, \dot{C}_\Lambda(x-y)\, \frac{\delta}{\delta \phi(x)}\left(P_{\Lambda}[\phi]\,\frac{\delta\Phi[\phi]}{\delta \phi(y)} \right)\,.
\end{align}
Integrating by parts twice in the functional $\phi$ derivatives, we obtain
\begin{align}
&-\int [d\phi] \int d^d x \,\frac{\delta}{\delta \phi(x)}\left(\int d^d y \, \frac{1}{2}\,\dot{C}_\Lambda(x-y) \, \frac{\delta \Sigma_\Lambda[\phi]}{\delta \phi(y)} \, P_{\Lambda}[\phi]\right) \Phi[\phi] \nonumber \\
& \qquad \qquad \qquad \qquad \qquad \qquad \qquad \qquad \qquad \quad = \int [d\phi] \int d^d x \, \frac{\delta}{\delta \phi(x)}\,(\Psi_\Lambda[\phi, x] \, P_\Lambda[\phi]) \, \Phi[\phi] \nonumber \\
& \qquad \qquad \qquad \qquad \qquad \qquad \qquad \qquad \qquad \quad = \left\langle -\int d^d x \, \frac{\delta}{\delta \phi(x)}\,(\Psi_\Lambda[\phi, x] \, P_\Lambda[\phi]), \,\delta P\right\rangle_{\mathcal{W}_2}
\end{align}
where in the last line we have used~\eqref{E:functionalRiemannian2}.  Comparing with~\eqref{E:functionalalgradient1} this establishes
\begin{equation}
- \nabla_{\mathcal{W}_2} S(P_\Lambda[\phi]\,\|\,Q_\Lambda[\phi]) = \int d^d x \, \frac{\delta}{\delta \phi(x)}\,(\Psi_\Lambda[\phi, x] \, P_\Lambda[\phi])
\end{equation}
which implies our main result~\eqref{E:boxed1}.

\subsection{Comments and interpretation}
\label{Subsec:commentsandinterp1}

Our result~\eqref{E:boxed1} provides a new way of thinking about the renormalization group, and elucidates some key technical aspects of Polchinski's equation and the Wegner-Morris flow equation more broadly.  First let us discuss~\eqref{E:boxed1} itself.

The relative entropy $S(P \| Q)$, also called the Kullback-Leibler divergence, is a core object in information theory which provides a measure of similarity between two probability distributions $P,Q$~\cite{cover1999elements}.  While it is not a metric distance (for instance, it is not symmetric between $P$ and $Q$ and does not satisfy the triangle inequality), it is positive and enjoys a host of other properties; a useful discussion aimed for physicists is~\cite{Witten:2018zva}.  Heuristically, the relative entropy tells us how good $Q$ is as a proxy for $P$.  For instance, the relative entropy quantifies how much additional memory is required to compress a list of samples from $P$ if we are only given just enough memory to optimally compress a list of as many samples from $Q$.  There have been other works on RG flow which have leveraged the relative entropy~\cite{Beny:2012qh, Beny:2014sna, Balasubramanian:2014bfa, lashkari2019entanglement, Furuya:2020tzv, Fowler:2021oje, Koenigstein:2021rxj, Erdmenger:2021sot}, albeit in a manner which does not involve optimal transport.

A particular conceptual feature of the relative entropy is worth commenting on.  If we have a discrete probability distribution $p_i$, then its entropy is simply $- \sum_i p_i \log(p_i)$.  Passing to the continuum via $p_i \to dx \, p(x)$, the entropy becomes $- \int dx \,p(x) \log(dx \, p(x))$.  The $dx$ inside the logarithm is somewhat pathological, and reflects that the strict continuum limit of the entropy is ill-defined. Relatedly, if we give $dx$ units of length so that $p(x)$ has units of inverse length, then the quantity inside the logarithm must be dimensionless, which is achieved by $\log(dx \, p(x))$. To cure the issue of a $dx$ inside the logarithm, the continuum entropy is obliged to have an alternative defining formula which is partially divorced from its discrete version.  A common option is $S[p] = - \int dx \, p(x) \log(p(x))$, which is called the differential entropy.  In the differential entropy, the $\log(p(x))$ should be thought of as $\log(a\,p(x))$ for $a = 1$, where $a$ has `units' of length.  Notably, the relative entropy is free of the aforementioned issue.  For suppose we consider $-\sum_i p_i \log(q_i)$ for some second discrete probability distribution $q_i$ and pass to the continuum limit in the same way to get $- \int dx \, p(x) \log(dx\,q(x))$.  Subtracting this from $- \int dx \,p(x) \log(dx \, p(x))$, we obtain minus the relative entropy
\begin{equation}
- S( p \| q) = - \int dx \, p(x) \log\left(\frac{p(x)}{q(x)}\right)
\end{equation}
where in effect the unwanted $dx$'s in the log have cancelled out.  As such, we can think of minus the relative entropy as a well-defined and meaningful replacement for the continuum entropy.

One interpretation of our result~\eqref{E:boxed1} is that the RG flow of the probability functional $P_\Lambda$ seeks to minimize the relative entropy between $P_\Lambda$ and $Q_\Lambda$ according to the appropriate $\mathcal{W}_2$ gradient.  Minimizing the relative entropy can be viewed as a proxy for maximizing the entropy of $P_\Lambda$, in light of the discussion in the preceding paragraph.  This makes intuitive sense: as we coarse-grain due to RG flow, there is a form of entropy production.  But what is more striking from~\eqref{E:boxed1} is that the entropy production is \textit{precisely} what determines the flow itself.  An alternative formulation of this statement is provided in Section~\ref{Sec:Formulations} where we develop a variational formula for RG flow.

An interesting special case of~\eqref{E:boxed1} is Polchinski's equation for a free scalar field, corresponding to
\begin{equation}
- \Lambda \frac{d}{d\Lambda} P_\Lambda^{\text{free}}[\phi] = - \nabla_{\mathcal{W}_2} S( P_\Lambda^{\text{free}}[\phi]\,\|\,Q_\Lambda^{\text{free}}[\phi])
\end{equation}
with $P_\Lambda^{\text{free}}[\phi] = e^{- S_{\text{free},\Lambda}[\phi]}/Z_{P,\Lambda}$ and  $Q_\Lambda^{\text{free}}[\phi] = e^{- 2 S_{\text{free},\Lambda}[\phi]}/Z_{Q,\Lambda}$ where we note the factor of $2$ in the exponent.  Using the identities
\begin{equation}
- \nabla_{\mathcal{W}_2} S( P_\Lambda^{\text{free}}[\phi]\,\|\,Q_\Lambda^{\text{free}}[\phi]) = - \nabla_{\mathcal{W}_2} S( P_\Lambda^{\text{free}}[\phi]\,\|\,(P_\Lambda^{\text{free}}[\phi])^2\,) = \nabla_{\mathcal{W}_2} S( P_\Lambda^{\text{free}}[\phi])
\end{equation}
where $S(P) = - \int [d\phi] \, P[\phi] \log P[\phi]$ is a functional analogue of the differential entropy, we find
\begin{equation}
- \Lambda \frac{d}{d\Lambda} P_\Lambda^{\text{free}}[\phi] =  \nabla_{\mathcal{W}_2} S( P_\Lambda^{\text{free}}[\phi])\,.
\end{equation}
Thus the free scalar field flows exactly according to its differential entropy.

Another feature of~\eqref{E:boxed1} is that it explains the role of `seed action' $\hat{S}_\Lambda[\phi]$ in~\eqref{E:Sigma1}.  In particular, the seed action (and its conventional prefactor of $2$) provides us with $Q_\Lambda[\phi] = e^{-2\hat{S}_\Lambda[\phi]}/Z_{Q,\Lambda}$ as per~\eqref{E:Q1}, which is the baseline distribution in the relative entropy $S(P_\Lambda \, \| \, Q_\Lambda)$ appearing in the gradient flow.  Indeed, the definition of the functional $\mathcal{W}_2$ distance together with $Q_\Lambda$ define our choice of RG scheme.  We emphasize that the seed action $\hat{S}_{\Lambda}[\phi]$ has a prescribed $\Lambda$-dependence, and does not itself need to satisfy a flow equation.

Finally, we comment on the meaning of $\Sigma_\Lambda[\phi]$.  Suggestively rewriting it as
\begin{equation}
\label{E:Sigma2}
\Sigma_\Lambda[\phi] = - \log(P_\Lambda[\phi]) + \log(Q_\Lambda[\phi]) - \log(Z_{P,\Lambda}) + \log(Z_{Q,\Lambda})\,,
\end{equation}
we observe that $\Sigma_\Lambda[\phi]$ ultimately enters into our formulas only through its functional derivative $\frac{\delta \Sigma_\Lambda[\phi]}{\delta \phi}$.  As such, we are free to redefine $\Sigma_\Lambda[\phi]$ by adding $\phi$-independent terms. Thus, subtracting the constant terms off of~\eqref{E:Sigma2} and combining the residual logarithms, we can replace $\Sigma_\Lambda[\phi]$ in~\eqref{E:Wegnerflow1},~\eqref{E:Psieq1} with 
\begin{equation}
\label{E:Sigma3}
\widetilde{\Sigma}_\Lambda[\phi] = - \log\left(\frac{P_\Lambda[\phi]}{Q_\Lambda[\phi]}\right)\,,
\end{equation}
where the tilde reminds us that we have made a modification (albeit an innocuous one) to the original definition without changing the resulting Wegner-Morris flow equation. Notice that this new quantity $\widetilde{\Sigma}_\Lambda[\phi]$ is information-theoretically natural: it is minus the log likelihood ratio between $P_\Lambda$ and $Q_\Lambda$, and so we can write
\begin{equation}
S(P_\Lambda \, \| \, Q_\Lambda) = - \int [d\phi]\, P_\Lambda[\phi] \, \widetilde{\Sigma}_\Lambda[\phi]\,.
\end{equation}
Accordingly, we have repackaged the major ingredients in the Wegner-Morris flow equation (and Polchinski's equation as a special case) in terms of information-theoretic quantities.

\section{RG monotones}
\label{Sec:Monotones}

In this section we derive a non-perturbative RG monotone using our optimal transport flow equation in~\eqref{E:boxed1}.  There have been previous attempts at formulating RG monotones using the ERG framework but this has only been successful in the local potential approximation (LPA), essentially where we ignore higher-derivative contributions to the action~\cite{Zumbach:1993zz, Zumbach:1994kc, Zumbach:1994vg, Generowicz:1997he}.  By contrast, our RG monotone holds without any approximations.

Our proposed monotone for a $P_\Lambda$ solving~\eqref{E:boxed1} is formally given by
\begin{equation}
\label{E:monotone1}
M_\Lambda(P_\Lambda) := S(P_\Lambda \, \|\, Q_\Lambda) - \log(Z_{Q,\Lambda})\,,
\end{equation}
under the assumption that $Q_\Lambda[\phi] = e^{- S_{Q,\Lambda}[\phi]}/Z_{Q,\Lambda}$ for
\begin{equation}
S_{Q,\Lambda}[\phi] = C \int \frac{d^d p}{(2\pi)^d} \, \widehat{K}_{\Lambda}^{-1}(p^2)\,G^{-1}(p^2) \, \phi(p) \phi(-p)\,.
\end{equation}
Here $\widehat{K}_{\Lambda}(p^2)$ is a smooth cutoff function which need not equal $K_\Lambda(p^2)$, and $G(p^2)$ is the Green's function of some positive semi-definite elliptic differential operator (e.g. $G(p^2) = 1/(p^2 + m^2)$).  Accordingly, our monotone pertains to Polchinski's equation, as well as more generally the Wegner-Morris flow equation with a quadratic seed action.

Due to interesting subtleties with orders of limits and divergences, in Section~\ref{subsec:regularize1} (see in particular~\eqref{E:truemonotoneboxed}) we will introduce a regulated version of $M_\Lambda(P_\Lambda)$.  We will show below that the quantity~\eqref{E:monotone1} is formally divergent, but can be regularized in a way that is independent of the renormalization scheme.  There is an extensive discussion in Section~\ref{subsec:regularize1} which provides appropriate context.  The proof of monotonicity below is unaffected.

\subsection{Proof of monotonicity}
\label{subsec:proof1}

Let us establish the monotonicity of the monotone.  We have
\begin{align}
\label{E:Sprimederiv1}
-\Lambda \frac{d}{d \Lambda} \, M_\Lambda(P_\Lambda) &= -\int [d\phi] \, \Lambda \frac{\partial P_\Lambda}{\partial \Lambda} \, (\log(P_\Lambda) - \log(e^{-S_{Q,\Lambda}})) - \int [d\phi] \,  \left(\Lambda \frac{\partial P_\Lambda}{\partial \Lambda} + P_\Lambda \, \frac{\partial S_{Q,\Lambda}}{\partial \Lambda}\right) \,.
\end{align}
Here we are differentiating under the integral sign by bringing $\Lambda \frac{d}{d\Lambda}$ into integrand of the functional integral.  This has some subtleties related to regularization of $M_\Lambda(P_\Lambda)$ which we treat in detail in Section~\ref{subsec:regularize1}, but indeed we will find that integrating under the integral sign is a good prescription.  Using~\eqref{E:boxed1} and dropping total derivative terms, we find
\begin{align}
\int [d\phi] \, \int d^d x\, \frac{\delta (\Psi_\Lambda P_\Lambda)}{\delta \phi(x)} \, (\log(P_\Lambda) - \log(e^{-S_{Q,\Lambda}})) - \int [d\phi] \,  P_\Lambda \, \Lambda \frac{\partial S_{Q,\Lambda}}{\partial \Lambda}\,.
\end{align}
Integrating by parts on the first term, we obtain
\begin{align}
\label{E:usefulperturbative1}
\frac{1}{2}\int [d\phi] \, P_\Lambda[\phi]\int d^d x \, d^d y\, \dot{C}_\Lambda(x-y) \, \frac{\delta \Sigma_\Lambda}{\delta \phi(x)} \frac{\delta \Sigma_\Lambda}{\delta \phi(y)}  - \int [d\phi] \,  P_\Lambda \, \Lambda \frac{\partial S_{Q,\Lambda}}{\partial \Lambda}\,.
\end{align}
The first term\footnote{It can also be written as functional generalization of the relative Fisher information between $P_\Lambda$ and $Q_\Lambda$ with background metric $\langle F[\phi(x)], G[\phi(y)] \rangle = \frac{1}{2}\int d^d x \, d^d y \, \dot{C}(x-y) \, F[\phi(x)] \,G[\phi(y)]$.} is manifestly positive semi-definite since $ \dot{C}_\Lambda(x-y) \geq 0$.  For the second term, we have
\begin{equation}
- \int [d\phi] \,  P_\Lambda \, \Lambda \frac{\partial S_{Q,\Lambda}}{\partial \Lambda} = -\int \frac{d^d p}{(2\pi)^d}\,\Lambda \frac{\partial \widehat{K}_{\Lambda}^{-1}(p^2)}{\partial \Lambda} \, G^{-1}(p^2) \, \langle \phi(p) \phi(-p)\rangle_{P_\Lambda}\,.
\end{equation}
Since $-\Lambda \frac{\partial \hat{K}^{-1}(p^2/\Lambda^2)}{\partial \Lambda} \geq 0$ and $\langle \phi(p) \phi(-p)\rangle_P \geq 0$, the entire quantity is greater than or equal to zero.  Accordingly, we have established that
\begin{equation}
\label{E:boxed2}
\boxed{- \Lambda \frac{d}{d \Lambda}\,M_\Lambda(P_\Lambda) \geq 0}
\end{equation}
and so $M_\Lambda(P_\Lambda)$ is an RG monotone.

A slight surprise about the definition of the monotone~\eqref{E:monotone1} is the presence of the $-\log(Z_{Q,\Lambda})$.  The necessity of this term can be understood as follows. Suppose we did not include $-\log(Z_{Q,\Lambda})$ in the monotone, so that $-\Lambda \frac{d}{d\Lambda} M_\Lambda(P_\Lambda)$ is just the relative entropy.  This would affect the left-hand side of~\eqref{E:boxed2} by adding the term
\begin{equation}
\label{E:extraterm1}
- \Lambda \frac{\partial\log(Z_{Q,\Lambda})}{\partial \Lambda} = - \Lambda \,\frac{1}{Z_{Q,\Lambda}} \frac{\partial Z_{Q,\Lambda}}{\partial \Lambda}\,.
\end{equation}
Unfortunately this term is less than zero, and so can interfere with the bound~\eqref{E:boxed2} if we include it.  In particular, as we raise the cutoff $\Lambda$, more modes are introduced in the $S_{Q,\Lambda}$ action, with variances $\sigma^2 \simeq 1/(p^2 + m^2)$ for $p \sim \Lambda$; the variances of these modes were formerly extremely small before we raised the cutoff.  Accordingly, the partition function $Z_{Q,\Lambda}$ will increase when we raise the cutoff, and so $\Lambda \frac{\partial Z_{Q,\Lambda}}{\partial \Lambda} \geq 0$.  This is why we have elected to define our monotone to avoid this issue.

Note, however, that if we had a hard cutoff instead of a soft cutoff, the story would be different.  In the hard cutoff setting, introducing more modes by raising the cutoff would cause $\Lambda \frac{\partial Z_Q}{\partial  \Lambda} \leq 0$, and render~\eqref{E:extraterm1} to be positive.  However, other subtleties with ERG in the hard cutoff setting pertaining to changing the domain of path integration dissuade us from pursuing this direction at present.

Having defined a non-perturbative RG monotone $M_\Lambda(P_\Lambda)$ for quantum field theories, it is natural to inquire about the finiteness of $M_\Lambda(P_\Lambda)$.  This will become clearer when we compute some examples below, but here we overview some general structure.  In our examples we will find that
\begin{equation}
\label{E:Mschematic1}
- \Lambda \frac{d}{d \Lambda}\,M_\Lambda(P_\Lambda) = \delta^d(0)\,c_1(\Lambda)
\end{equation}
where $c_1(\Lambda)$ is a finite quantity.  Here the $\delta^d(0)$ divergence comes from momentum space contact terms.  If we considered a field theory on, say, a torus where the momenta range over a lattice, then the $\delta^d(0)$ would be rendered finite.
Since the $\delta^d(0)$ is multiplicative on the right-hand side of~\eqref{E:Mschematic1} it is essentially innocous: the positivity of $- \Lambda \frac{d}{d \Lambda}\,M_\Lambda(P_\Lambda)$ implies
\begin{equation}
c_1(\Lambda) \geq 0\,.
\end{equation}
Then an appropriate anti-derivative of $c_1(\Lambda)$, namely a $C_1(\Lambda)$ satisfying $- \Lambda \frac{d}{d\Lambda} \,C_1(\Lambda) = c_1(\Lambda)$, is evidently a finite RG monotone since
\begin{equation}
- \Lambda \frac{d}{d\Lambda} \,C_1(\Lambda) \geq 0\,.
\end{equation}

Our discussion above pertained to $- \Lambda \frac{d}{d \Lambda}\,M_\Lambda(P_\Lambda)$ instead of $M_\Lambda(P_\Lambda)$ itself.  As we will see in in Section~\ref{subsec:regularize1} below, there are some subtleties in computing $M_\Lambda(P_\Lambda)$ directly.  It can be done, however, with sufficient care.  Nonetheless, the derivative $- \Lambda \frac{d}{d \Lambda}\,M_\Lambda(P_\Lambda)$ can be computed rather directly using the formula~\eqref{E:usefulperturbative1} which automatically accounts for subtleties in the definition of $M_\Lambda(P_\Lambda)$.

\subsection{Differentiating under the functional integral and regularization}
\label{subsec:regularize1}

In our derivation of the monotonicity of $M_\Lambda(P_\Lambda)$ above, we differentiated under the integral sign in~\eqref{E:Sprimederiv1}.  This interchange of limits is particularly subtle in our path integral setting as we will now show.  Let us start with an illuminating example before turning to generalities.

\subsubsection{Order of limits in the setting of free field theory}
Consider Polchinski's equation in~\eqref{E:convectiondiffusion1}; a solution to this is the free probability distribution, given by $P_\Lambda[\phi] = \frac{1}{Z_{\Lambda}}\,\exp\left(-\frac{1}{2}\int \frac{d^d p}{(2\pi)^d} \, \phi(p)\phi(-p)(p^2 + m^2) K_\Lambda^{-1}(p^2)\right)$.  Plugging this into our monotone $M_\Lambda(P_\Lambda)$ defined in~\eqref{E:monotone1}, there are terms proportional to
\begin{align}
\label{E:constantdivergence1}
&\int [d\phi]\, P_\Lambda^{\text{free}}[\phi] \, S_{\text{free}, \Lambda}[\phi] \nonumber \\
& \qquad \qquad = \frac{1}{2 Z_{P,\Lambda}} \!\int [d\phi]\,e^{- \frac{1}{2}\int \!\! \frac{d^d p}{(2\pi)^d} \, \phi(p)\phi(-p)(p^2 + m^2) K_\Lambda^{-1}(p^2)} \!\!\! \int \!\!\frac{d^d p}{(2\pi)^d} \, \phi(p)\phi(-p)(p^2 + m^2) K_\Lambda^{-1}(p^2) \nonumber \\
& \qquad \qquad =  \frac{1}{2}\int d^d p \, \delta^d(0)\,.
\end{align}
This is related to the fact that in the more ordinary $n$-dimensional integral setting we have $\int d^n x \, \frac{\det(A)^{1/2}}{(2\pi)^{n/2}}\, e^{- \frac{1}{2}\, x \cdot A \cdot x} \left( \frac{1}{2}\, x \cdot A \cdot x \right)= \frac{1}{2}$\,, except that in the functional setting we have a (momentum space) contact term $\delta^d(0)$ and a residual $\int d^d p$ integral at the end.  From~\eqref{E:constantdivergence1} we infer that
\begin{equation}
\label{E:firstway11}
- \Lambda \frac{d}{d\Lambda}\int [d\phi]\, P_\Lambda[\phi] \, S_{\text{free},\Lambda}[\phi] = 0\,.
\end{equation}

But now let us perform the computation of~\eqref{E:firstway11} another way, by differentiating under the integral sign.  In this setting we have
\begin{align}
\label{E:secondway11}
- \int [d\phi]\, \Lambda \frac{\partial}{\partial \Lambda}\left(P_\Lambda^{\text{free}}[\phi] \, S_{\text{free},\Lambda}[\phi]\right) &= - \int [d\phi]\, \Lambda \frac{\partial P_\Lambda^{\text{free}}[\phi]}{\partial \Lambda} \, S_{\text{free},\Lambda}[\phi] - \int [d\phi]\,P_\Lambda^{\text{free}}[\phi]\,\Lambda \frac{\partial S_{\text{free},\Lambda}[\phi]}{\partial \Lambda}\,.
\end{align}
The first term on the right-hand side is
\begin{align}
\label{E:firsttermalot1}
&- \int [d\phi]\, \Lambda \frac{\partial P_\Lambda^{\text{free}}[\phi]}{\partial \Lambda} \, S_{\text{free},\Lambda}[\phi] \nonumber \\ \nonumber \\
& \qquad =  \frac{1}{4 Z_{P,\Lambda}} \!\int [d\phi]\,e^{- \frac{1}{2}\int \!\! \frac{d^d p}{(2\pi)^d} \, \phi(p)\phi(-p)(p^2 + m^2) K_\Lambda^{-1}(p^2)} \!\!\! \int \!\!\frac{d^d p}{(2\pi)^d} \, \phi(p)\phi(-p)(p^2 + m^2) K_\Lambda^{-1}(p^2) \nonumber \\
&\qquad \qquad \qquad \qquad \qquad \qquad \qquad \qquad \qquad \qquad \qquad \quad \times \int \!\!\frac{d^d q}{(2\pi)^d} \, \phi(q)\phi(-q)(q^2 + m^2) \Lambda \frac{\partial K_\Lambda^{-1}(q^2)}{\partial \Lambda} \nonumber \\
& \qquad \qquad + \Lambda \, \frac{\partial \log Z_{P,\Lambda}}{\partial \Lambda} \int [d\phi] \, P_\Lambda^{\text{free}}[\phi] \, S_{\text{free},\Lambda}[\phi] \nonumber \\
& \qquad = \left(\frac{1}{2}\!\int d^d p \, \delta^d(0) \! \right) \!\! \left(\Lambda  \frac{\partial \log Z_{P,\Lambda}}{\partial \Lambda}\! -\! \frac{1}{2}\delta^d(0) \!\int d^d p \,\Lambda \frac{\partial \log K_{\Lambda}(p^2)}{\partial \Lambda} \right)\! - \frac{1}{2}  \delta^d(0)\! \int d^d p \, \Lambda \frac{\partial \log K_{\Lambda}(p^2)}{\partial \Lambda}\,.
\end{align}
To further simplify, we observe that for an infinitesimal perturbation change of scale any $P_{\Lambda}[\phi]$ changes as $P_{\Lambda - \delta \Lambda}[\phi] = P_\Lambda[\phi] - \delta \Lambda \, \frac{\partial P_\Lambda[\phi]}{\partial \Lambda}$.  Then integrating both sides with respect to $\phi$ and using the normalization of the probability functional we find 
\begin{equation}
\label{E:usefulfree1}
\int [d\phi] \, \Lambda \frac{\partial P_\Lambda[\phi]}{\partial \Lambda} = 0\,.
\end{equation}
In the free setting,
\begin{equation}
\int [d\phi] \, \Lambda \frac{\partial P_\Lambda^{\text{free}}[\phi]}{\partial \Lambda} = \frac{1}{2}\,\delta^d(0)\int d^d p \, \Lambda \frac{\partial \log K_\Lambda(p^2)}{\partial \Lambda} - \Lambda \frac{\partial \log Z_{P,\Lambda}}{\partial \Lambda} = 0
\end{equation}
which implies
\begin{equation}
\label{E:usefulfree2}
\Lambda \frac{\partial \log Z_{P,\Lambda}}{\partial \Lambda} = \frac{1}{2}\,\delta^d(0)\int d^d p \, \Lambda \frac{\partial \log K_\Lambda(p^2)}{\partial \Lambda}\,.
\end{equation}
Plugging this into~\eqref{E:firsttermalot1} we see that there is a helpful cancellation which leaves us with
\begin{equation}
\label{E:helpful1}
- \int [d\phi]\, \Lambda \frac{\partial P_\Lambda^{\text{free}}[\phi]}{\partial \Lambda} \, S_{\text{free},\Lambda}[\phi] = - \frac{1}{2}\,\delta^d(0)\int d^d p \, \Lambda \frac{\partial \log K_\Lambda(p^2)}{\partial \Lambda}\,.
\end{equation}
Turning to the second term on the right-hand side of~\eqref{E:secondway11}, a less elaborate computation yields
\begin{equation}
\label{E:helpful2}
- \int [d\phi]\,P_\Lambda^{\text{free}}[\phi]\,\Lambda \frac{\partial S_{\text{free},\Lambda}[\phi]}{\partial \Lambda} = \delta^d(0)\int d^d p \, \Lambda \frac{\partial \log K_\Lambda(p^2)}{\partial \Lambda}\,.
\end{equation}
Plugging~\eqref{E:helpful1} and~\eqref{E:helpful2} into~\eqref{E:secondway11}, we finally find
\begin{equation}
\label{E:secondwayanswer1}
- \int [d\phi]\, \Lambda \frac{\partial}{\partial\Lambda}\left(P_\Lambda[\phi] \, S_{\text{free},\Lambda}[\phi]\right) = \frac{1}{2}\,\delta^d(0)\int d^d p \, \Lambda \frac{\partial \log K_\Lambda(p^2)}{\partial \Lambda}\,.
\end{equation}

Comparing~\eqref{E:secondwayanswer1} with~\eqref{E:firstway11}, we see that surprisingly
\begin{equation}
\label{E:notcommuting1}
- \Lambda \frac{d}{d\Lambda} \int [d\phi]\, P_\Lambda[\phi] \, S_{\text{free},\Lambda}[\phi]\not = - \int [d\phi]\, \Lambda \frac{\partial}{\partial\Lambda}\left(P_\Lambda[\phi] \, S_{\text{free},\Lambda}[\phi]\right)\,,
\end{equation}
and so evidently the order of limits does not commute.  This also holds more generally for RG flows of interacting theories.  In our proof of the monotonicity of $M_\Lambda(P_\Lambda)$, we differentiated under the integral sign, and so apparently our proof is contingent on a certain order of limits.  To resolve the ambiguity, let us \textit{define}
\begin{equation}
\label{E:truemonotone1}
M_\Lambda(P_{\Lambda}^{\text{free}}[\phi]) :=  - \int_{\log \Lambda}^{\log \Lambda_0} d\log\Lambda' \left[\int [d\phi] \,\Lambda' \frac{\partial}{\partial \Lambda'} \left( P_{\Lambda'}^{\text{free}}[\phi] \log\left(\frac{P_{\Lambda'}^{\text{free}}[\phi]}{Q_{\Lambda'}^{\text{free}}[\phi]}\right)\right) - \Lambda' \frac{\partial \log(Z_{Q,\Lambda'})}{\partial \Lambda'} \right]
\end{equation}
where we assume $\Lambda \leq \Lambda_0$.  In words, we are defining $M_\Lambda(P_{\text{free},\Lambda}[\phi])$ as an anti-derivative of the differentiated-under-the-integral-sign quantity.  This more fully specifies what we mean by $M_\Lambda(P_{\text{free},\Lambda}[\phi])$, and in particular the manner in which its divergent terms depend on $\Lambda$.

\subsubsection{Order of limits in more general RG flows}

In the general setting, by analogy to~\eqref{E:truemonotone1} we define
\begin{equation}
\label{E:truemonotoneboxed}
\boxed{M_\Lambda(P_{\Lambda}[\phi]) :=  - \int_{\log \Lambda}^{\log \Lambda_0} d\log\Lambda' \left[\int [d\phi] \,\Lambda' \frac{\partial}{\partial \Lambda'} \left( P_{\Lambda'}[\phi] \log\left(\frac{P_{\Lambda'}[\phi]}{Q_{\Lambda'}[\phi]}\right)\right) - \Lambda' \frac{\partial \log(Z_{Q,\Lambda'})}{\partial \Lambda'} \right]}
\end{equation}
where again we assume $\Lambda \leq \Lambda_0$.  This is the true definition of the monotone $M_\Lambda(P_\Lambda)$.  Indeed, the proof of monotonicity in Section~\ref{subsec:proof1} in fact implicitly uses this prescription.

We conclude this section by reiterating a useful formula we used in our free analysis above.  Equation~\eqref{E:usefulfree1} is $\int [d\phi] \, \Lambda \frac{\partial P_\Lambda[\phi]}{\partial \Lambda} = 0$ which holds for general $P_\Lambda[\phi]$, and so 
\begin{equation}
\label{E:usefulgeneral1}
\Lambda \frac{\partial \log Z_{P,\Lambda}}{\partial \Lambda} = -\int [d\phi] \, P_\Lambda[\phi] \, \Lambda \frac{\partial S_{P,\Lambda}[\phi]}{\partial \Lambda}
\end{equation}
which is just a generalization of~\eqref{E:usefulfree2}.  This identity~\eqref{E:usefulgeneral1} will be useful for us in the section which follows.

\section{Examples with scalar field theories}
\label{Sec:Examples}

Below we exhibit computations of our RG monotone in some examples.  First we consider free scalar field theory which has an exactly soluble ERG flow; hence we can compute our RG monotone exactly in this case.  Next we turn to scalar $\phi^4$ theory for which proceed perturbatively.

\subsection{Free scalar field}

Consider a free massive scalar field which evolves via Polchinski's equation in~\eqref{E:convectiondiffusion1}.  Using the definition of $M_\Lambda(P_{\Lambda}^{\text{free}}[\phi])$ in~\eqref{E:truemonotone1}, we find
\begin{equation}
M_\Lambda(P_{\Lambda}^{\text{free}}[\phi]) = - \frac{3}{2} \, \delta^d(0) \int d^d p \, \log\!\left(\frac{K_\Lambda(p^2)}{K_{\Lambda_0}(p^2)}\right)\,,
\end{equation}
where we suppose $\Lambda \leq \Lambda_0$\,.  The integral $\int d^d p \, \log\!\left(\frac{K_\Lambda(p^2)}{K_{\Lambda_0}(p^2)}\right)$ above is divergent, but its $\Lambda$ derivatives can be finite.  Upon taking a $\Lambda$ derivative, we find
\begin{equation}
- \Lambda \frac{d}{d\Lambda} \,M_\Lambda(P_{\Lambda}^{\text{free}}[\phi]) = \frac{3}{2}\,\delta^d(0) \int d^d p \, \frac{\partial \log K_{\Lambda}(p^2)}{\partial \Lambda} \geq 0\,.
\end{equation}
Although the above is positive for any $K_{\Lambda}(p^2)$ that is a smooth, monotonically decreasing cutoff function, it can be infinite.  There is a nice class of $K_{\Lambda}(p^2)$ for which the above is finite; this is explained in Appendix~\ref{App:convenient}.

\subsection{Interacting scalar field}

Now we perform some explicit perturbative computations of the derivative of the RG monotone $M_\Lambda(P_\Lambda)$ in~\eqref{E:truemonotoneboxed} for massive scalar $\phi^4$ theory, namely where the action is
\begin{align}
S_{\Lambda = \Lambda_0}[\phi] &= \frac{1}{2}\int \frac{d^d p}{(2\pi)^d}\,(p^2 + m^2)\,K^{-1}(p^2/\Lambda_0^2)\, \phi(p) \phi(-p) \nonumber \\
& \qquad + \frac{\lambda}{4!} \int \frac{d^d p_1 \, d^d p_2 \, d^d p_3 \, d^d p_4}{(2\pi)^{3d}}\, \phi(p_1) \phi(p_2) \phi(p_3)\phi(p_4)\,\delta^d(p_1 + p_2 + p_3 + p_4)\,.
\end{align}
We have given the action at the initial value of the cutoff $\Lambda = \Lambda_0$, where the RG flow is to be initiated.  In other words, we desire to study the flow equation for $P_\Lambda[\phi]$ given its initial condition at $\Lambda = \Lambda_0$.  Recall that in the context of Polchinski's equation we have
\begin{align}
\label{E:ourkernels1}
\dot{C}_\Lambda(p^2) = (2\pi)^d (p^2 + m^2)^{-1} \,\Lambda \frac{\partial K_\Lambda(p^2)}{\partial \Lambda}\,,\quad 
\hat{S}_\Lambda = \frac{1}{2}\int \frac{d^d p}{(2\pi)^d}\,(p^2 + m^2)\, K_\Lambda^{-1}(p^2)\,\phi(p) \phi(-p)\,.
\end{align}

Equation~\eqref{E:usefulperturbative1} provides a nice expression for the derivative of our RG monotone at $\Lambda = \Lambda_0$, namely
\begin{align}
&- \Lambda \frac{d}{d\Lambda}\,M_\Lambda(P_\Lambda)\Big|_{\Lambda = \Lambda_0} \nonumber \\
& \qquad \quad = \frac{1}{2}\int [d\phi] \, P_{\Lambda_0}[\phi]\int d^d x \, d^d y\, \dot{C}_{\Lambda_0}(x-y) \, \frac{\delta \Sigma_{\Lambda_0}}{\delta \phi(x)} \frac{\delta \Sigma_{\Lambda_0}}{\delta \phi(y)}  - 2 \int [d\phi] \,  P_{\Lambda_0}[\phi] \, \Lambda_0 \frac{\partial \hat{S}_{\Lambda_0}}{\partial \Lambda_0}\,.
\end{align}
Working in momentum space and plugging in~\eqref{E:ourkernels1} we find
\begin{align}
\label{E:tocompute1}
&\frac{1}{2}\int [d\phi] \, P_{\Lambda_0}[\phi]\int d^d p\,(2\pi)^d (p^2 + m^2)^{-1} \,\Lambda_0 \frac{\partial K(p^2/\Lambda_0^2)}{\partial \Lambda_0}\, \frac{\delta (S_{\Lambda_0} - 2\hat{S}_{\Lambda_0})}{\delta \phi(p)} \frac{\delta (S_{\Lambda_0} - 2\hat{S}_{\Lambda_0})}{\delta \phi(-p)}  \nonumber \\
& \qquad \qquad \qquad \qquad \qquad \qquad \qquad \qquad \qquad \qquad \qquad \qquad \qquad \quad  - 2 \int [d\phi] \,  P_{\Lambda_0}[\phi] \, \Lambda_0 \frac{\partial \hat{S}_{\Lambda_0}}{\partial \Lambda_0}\,.
\end{align}
We will compute this perturbatively to second order in the quartic coupling $\lambda$.

To compress the form of our formulas, it is convenient to define
\begin{equation}
\widetilde{K}_\Lambda(p) := K_\Lambda(p^2) \, \frac{1}{p^2 + m^2}
\end{equation}
and also
\begin{align}
f_0(\Lambda) &= \int d^d p \,\Lambda \, \frac{\partial \log(\widetilde{K}_{\Lambda}(p))}{\partial \Lambda} \\
f_1(\Lambda) &= \int d^d p \, \Lambda \frac{\partial \log(\widetilde{K}_{\Lambda}(p))}{\partial \Lambda}\, \widetilde{K}_{\Lambda}(p) \int \frac{d^d q}{(2\pi)^d}\,\widetilde{K}_{\Lambda}(q) \\
f_2(\Lambda) &= \int  d^d p \,\Lambda \, \frac{\partial \log(\widetilde{K}_{\Lambda}(p))}{\partial \Lambda} \, \widetilde{K}_{\Lambda}(p)  \int \frac{d^d q_1}{(2\pi)^d} \frac{d^d q_2}{(2\pi)^d}\, \widetilde{K}_{\Lambda}(p\!-\!q_1\!-\!q_2) \widetilde{K}_{\Lambda}(q_1) \widetilde{K}_{\Lambda}(q_2) 
\end{align}
which are all finite for appropriate choices of $K_\Lambda(p^2)$ (see Appendix~\ref{App:convenient}).
Above, we have written the $\frac{\partial \log(\widetilde{K}_{\Lambda}(p))}{\partial \Lambda}$ term to make all the $f_i(\Lambda)$'s have the same form, but it can also be clarifying to simplify $f_1(\Lambda)$ and $f_2(\Lambda)$ using the identity $\frac{\partial \log(\widetilde{K}_{\Lambda}(p))}{\partial \Lambda} \, \widetilde{K}_\Lambda(p) = \frac{\partial \widetilde{K}_{\Lambda}(p)}{\partial \Lambda}$.  With our notation at hand, the main quantities in~\eqref{E:tocompute1} are
\begin{align}
\left\langle -2 \Lambda_0 \frac{\partial \hat{S}_{\Lambda_0} }{\partial \Lambda_0} \right\rangle_{P_{\Lambda_0}} &= \delta^d(0)\left(f_0(\Lambda_0) + \frac{\lambda^2}{6}\, f_2(\Lambda_0) + O(\lambda^3)\right) \\
\left\langle \! 2 \!\int\!d^d p \, (2\pi)^d \Lambda_0 \frac{\partial \widetilde{K}_{\Lambda_0}}{\partial \Lambda_0} \frac{\delta \hat{S}_{\Lambda_0}}{\delta \phi(p)} \frac{\delta \hat{S}_{\Lambda_0}}{\delta \phi(-p)} \!\right\rangle_{P_{\Lambda_0}} &= \delta^d(0)\left(2  f_0(\Lambda_0) +  \frac{\lambda^2}{3}\, f_2(\Lambda_0) + O(\lambda^3)\right) \\
\left\langle \!- 2\!\int\! d^d p \, (2\pi)^d \Lambda_0 \frac{\partial \widetilde{K}_{\Lambda_0}}{\partial \Lambda_0} \frac{\delta S_{\Lambda_0}}{\delta \phi(p)} \frac{\delta \hat{S}_{\Lambda_0}}{\delta \phi(-p)} \!\right\rangle_{P_{\Lambda_0}} &= \delta^d(0)\left(-2\, f_0(\Lambda_0) - \lambda \, f_1(\Lambda_0) -  \frac{2\lambda^2}{3}\, f_2(\Lambda_0) + O(\lambda^3)\right)  \\
\left\langle \frac{1}{2}\!\int\! d^d p \, (2\pi)^d \Lambda_0 \frac{\partial \widetilde{K}_{\Lambda_0}}{\partial \Lambda_0}\frac{\delta S_{\Lambda_0}}{\delta \phi(p)} \frac{\delta S_{\Lambda_0}}{\delta \phi(-p)} \!\right\rangle_{P_{\Lambda_0}} &= \delta^d(0)\left(\frac{1}{2} \, f_0(\Lambda_0) + \frac{\lambda}{2} \, f_1(\Lambda_0) + \frac{11 \lambda^2}{24}\, f_2(\Lambda_0) + O(\lambda^3)\right)  \,.
\end{align}
Plugging these into~\eqref{E:tocompute1} and simplifying (i.e., we just add up the above four equations), we find 
\begin{align}
- \Lambda \frac{d}{d\Lambda}\,M_\Lambda(P_\Lambda)\Big|_{\Lambda = \Lambda_0} &= \delta^d(0)\Bigg(\frac{3}{2}\,f_0(\Lambda_0) - \frac{1}{2}\,\lambda \, f_1(\Lambda_0) + \frac{7}{24}\, \lambda^2\, f_2(\Lambda_0) + O(\lambda^3)\Bigg)\,.
\end{align}
For perturbatively small $\lambda$, the above is greater than or equal to zero as it ought to be.

Our above computation shows how the RG monotone changes along an infinitesimal step of the flow, $\Lambda_0 \to \Lambda_0 - \delta \Lambda$.  We could continue with the next perturbative step along the flow, corresponding to computing $- \Lambda \frac{d}{d\Lambda}\,M_\Lambda(P_\Lambda)\big|_{\Lambda = \Lambda_0 - \delta \Lambda}$\,.  Thereafter we could continue on from there to successively smaller cutoff scales, but we will not pursue this here.

\section{Variational formulation of RG flows}
\label{Sec:Formulations}

We have established that Wegner-Morris flow is equivalent to the gradient flow of relative entropy with respect to a Wasserstein-$2$ distance on the space of fields.  In this section, we show that this connection allows us to construct a variational formulation of RG flow, which may be amenable to numerical methods.  Moreover, our analysis here establishes a new and precise connection between RG flows of conventional quantum field theories and numerical methods based on neural networks, which has previously only been established on a heuristic level. 

\subsection{Variational discretization of the renormalization group flow}

Consider $\mathbb{R}^n$ as a Riemannian manifold with the Euclidean metric, and let $F$ be a differentialble function $F : \mathbb{R}^n \to \mathbb{R}$.  Then a solution to the gradient flow equation
\begin{equation}
\label{E:gradflowsimple1}
\frac{d}{dt}\, X(t) = - \nabla F(X(t))
\end{equation}
with $X(0) = X_0$ can be approximated by a sequence of elements $X_0, X_\tau, X_{2\tau},...$ solving
\begin{equation}
\label{E:discrete-gradient-flow}
\frac{X_{(n+1)\tau} - X_{n \tau}}{\tau}  = - \Grad F(X_{(n+1)\tau})\,.
\end{equation}
For smaller $\tau$, the approximation becomes better.  We can equivalently recast~\eqref{E:discrete-gradient-flow} as an optimization problem,
\begin{equation}
\label{E:discrete-gradient-flow-0}
 X_{(n+1) \tau} = \argmin_X \left(\frac{1}{2 \tau}\,|X - X_{n \tau}|^2 + F(X)\right)\,.
\end{equation}

These considerations also apply to a more general Riemannian manifold $\mathcal{M}$ and a function $F: \mathcal{M} \to \R$.  In this setting a candidate approximate solution to~\eqref{E:gradflowsimple1} with $X(0) = X_0$ is given by a sequence of elements $X_0, X_\tau, X_{2\tau},...$ solving
\begin{equation}
\label{eq:discrete-gradient-flow0}
    X_{(n+1)\tau} = \argmin_X \left( \frac{1}{2 \tau}\,d(X, X_{n\tau})^2 + F(X)\right)\,,
\end{equation}
where $d(x,y)$ is the distance function on $\mathcal{M}$. This \emph{implicit Euler discretization scheme} can be proven in many cases~\cite{ambrosio2005gradient} to give approximate solutions that converge to the solution to the gradient flow equation~\eqref{E:gradflowsimple1} in the following sense. 
For any fixed time $T > 0$, if we let $X(t)$ be the unique solution to the gradient flow equation~\eqref{E:gradflowsimple1} with $X(0) = X_0$, then as $\tau \to 0$ we have 
\begin{equation}
    \textsf{Error}(\tau) = \sup_{n =0, \ldots, \lfloor T/\tau \rfloor } d(X_{n \tau}, X(n\tau)) \,\,\longrightarrow\,\, 0\,.
\end{equation}
In this manner, we can often approximate a gradient flow on a Riemannian manifold by a solution to a sequence of optimization problems.
The discussion in~\cite[Chapter 8.4]{villani2003topics} sketches the general scheme of proofs for such results; a careful general discussion of several cases of this convergence result, e.g.~for geodesically convex functionals $F$ on non-positively-curved manifolds $\M$, can be found in~\cite{ambrosio2005gradient}. 

Since we have recasted RG flow as a gradient flow of relative entropy with respect to a Wasserstein-2 metric, it is natural to ask if there is an approximation to RG flow along the lines of~\eqref{eq:discrete-gradient-flow0}.  We will find that indeed there is such an approximation, and that it is amenable to numerical optimization methods.  In the finite-dimensional context, implicit gradient numerical methods, now called JKO schemes, which simulate partial differential equations arising from gradient flows of entropy-like functionals, were first proposed by the pioneering~\cite{JKO}. In particular,~\cite{JKO} proves that the implicit Euler scheme~\eqref{eq:discrete-gradient-flow0} in the setting of the gradient flow of the entropy on Wasserstein space converges to the heat equation, and thus establishes the validity of this scheme in the finite-dimensional analogue of the setting of statistical field theory. For large gradient steps $\tau$, these methods require an efficient algorithmic approximation of the Wasserstein distance, which is available via the Sinkhorn algorithm~\cite{sinkhorn}. Numerical methods based on the JKO scheme are a topic of current interest in the applied mathematics community~\cite{wasserstein_gradient_applied_1, wasserstein_gradient_applied_2, wasserstein_gradient_applied_3, wasserstein_gradient_applied_4, wasserstein_gradient_applied_5, wasserstein_gradient_applied_6}, and in particular a number of recent proposals are based on approximating the Wasserstein distance by a neural-network-based method~\cite{wasserstein_gradient_neural_1, wasserstein_gradient_neural_2, wasserstein_gradient_neural_3}, analogous to the methodology that we propose here. Below, we explain the basic variational equations arising from the discretization of RG flows, and propose a novel numerical algorithm to compute the flow.

Recall from~\eqref{E:field-reparametrization} that Wegner-Morris flow is in fact a field reparameterization.  Suppose our initial probability functional is $P_{\Lambda_0}[\phi]$ at some scale $\Lambda_0$, and that we want to flow it to $P_{\Lambda_0 - t}[\phi]$.  Then the Wegner-Morris equation says that this can be expressed as
\begin{equation}
P_{\Lambda_0 - t}[\phi] = \left|\frac{\delta \mathcal{R}_{t}[\phi]}{\delta \phi}\right| \, P_{\Lambda_0}[\mathcal{R}_{t}[\phi]]
\end{equation}
for some reparameterization $\mathcal{R}_{t}$ which takes fields to fields.  Note that $\left|\frac{\delta \mathcal{R}_{t}[\phi]}{\delta \phi}\right| P_{\Lambda_0}[\mathcal{R}_{t}[\phi]] = \mathcal{R}_{t\,*}^{-1} P_{\Lambda_0}$ where $\mathcal{R}_{t\,*}^{-1} P_{\Lambda_0}$
is the pushforward of $P_{\Lambda_0}$ by the compositional inverse $\mathcal{R}_{t}^{-1}$
of $\mathcal{R}_t$\,.

For ease of notation, let us define
\begin{equation}
P_{\Lambda_0}^{\mathcal{R}}[\phi]:= \left|\frac{\delta \mathcal{R}[\phi]}{\delta \phi}\right| \, P_{\Lambda_0}[\mathcal{R}[\phi]]
\end{equation}
for an arbitrary reparameterization map $\mathcal{R}$.  Then we claim that a solution $P_\Lambda[\phi]$ to
\begin{equation*}
- \Lambda \frac{d}{d\Lambda} P_\Lambda[\phi] = - \nabla_{\mathcal{W}_2} S( P_\Lambda[\phi]\,\|\,Q_\Lambda[\phi])
\end{equation*}
with initial probability functional $P_{\Lambda_0}[\phi]$ satisfies
\begin{equation}
\label{E:funcapprox0}
P_{\Lambda_0 - \tau}[\phi] \approx P_{\Lambda_0}^{\mathcal{R}_{\tau}}[\phi]
\end{equation}
for small $\tau$, where
\begin{equation}
\label{E:variational0}
\boxed{\mathcal{R}_{\tau} = \argmin_{\mathcal{R}} \left( \frac{1}{2 \tau}\,\mathcal{W}_2(P_{\Lambda_0}^{\mathcal{R}}, P_{\Lambda_0})^2 + S(P_{\Lambda_0}^{\mathcal{R}} \, \| \, Q_{\Lambda_0})\right)}
\end{equation}
More generally, consider a sequence of reparameterizations $\mathcal{R}_\tau, \mathcal{R}_{2\tau}, \mathcal{R}_{3\tau},...$ and define
\begin{equation}
\label{E:Rtilde1}
\widetilde{\mathcal{R}}_{n \tau}:= \mathcal{R}_{n \tau}\circ \cdots \circ \mathcal{R}_{2\tau}\circ \mathcal{R}_\tau\,.
\end{equation}
Then we have
\begin{equation}
\label{E:funcapprox1}
P_{\Lambda_0 - n \tau}[\phi] \approx P_{\Lambda_0}^{\widetilde{\mathcal{R}}_{n\tau}}[\phi]
\end{equation}
if the $\mathcal{R}_n$'s satisfy
\begin{equation}
\label{E:variational1}
\boxed{\mathcal{R}_{(n+1)\tau} = \argmin_{\mathcal{R}} \left( \frac{1}{2 \tau}\,\mathcal{W}_2(P_{\Lambda_0}^{\mathcal{R}}, P_{\Lambda_0}^{\widetilde{\mathcal{R}}_{n\tau}})^2 + S(P_{\Lambda_0}^{\mathcal{R}} \, \| \, Q_{\Lambda_0 - n \tau})\right)}
\end{equation}
Moreover, the approximation~\eqref{E:funcapprox1} should become exact as $\tau \to 0$, by analogy to the finite dimensional JKO scheme~\cite{JKO}. We emphasize that~\eqref{E:variational1} is striking since it provides a variational formulation of RG flow. 

A natural question is if~\eqref{E:variational1} defines the $\mathcal{R}_{n\tau}$'s uniquely.  For simplicity, let us consider the $n = 1$ case, given in~\eqref{E:variational0}.  A solution to~\eqref{E:variational0} is supposed to provide us with an $\mathcal{R}_\tau$ such that $P_{\Lambda_0 - \tau}[\phi] \approx P_{\Lambda_0}^{\mathcal{R}_\tau}[\phi]$, becoming exact in the $\tau \to 0$ limit.  However, there exist many reparameterizations $\mathcal{R}_\tau'$ such that
\begin{equation}
P_{\Lambda_0}^{\mathcal{R}_\tau'}[\phi] = P_{\Lambda_0}^{\mathcal{R}_\tau}[\phi]\,.
\end{equation}
The fact at play here is that given a fixed probability distribution, there are many reparameterizations which transform that distribution in the same way.  Accordingly, a solution to~\eqref{E:variational0} is not unique, nor are solutions to~\eqref{E:variational1}.  While this non-uniqueness may seem bothersome, we will see shortly that the flexibility it provides is a virtue.

Examining the variational formulation of RG flow in~\eqref{E:variational1}, an undesirable aspect in practice is that the $\mathcal{W}_2(P_{\Lambda_0}^{\mathcal{R}}, P_{\Lambda_0}^{\widetilde{\mathcal{R}}_{n\tau}})^2$ term itself requires an optimization to compute, on account of the infimum in~\eqref{E:functionaltotalW2}.  However, we can get rid of this infimum in the following, interesting way.

We begin by considering~\eqref{E:variational0} as the $n = 1$ case of~\eqref{E:variational1}.  Recasting $\mathcal{W}_2(P_{\Lambda_0}^{\mathcal{R}}, P_{\Lambda_0})^2$ in the Monge formulation (i.e.~with the plausible assumption that our Kantorovich solutions are also Monge solutions), we find
\begin{align}
\label{E:newinf0}
&\mathcal{W}_2(P_{\Lambda_0}^{\mathcal{R}}, P_{\Lambda_0})^2 \nonumber \\
& \qquad = \inf_{\left\{\mathcal{F} \,:\, \mathcal{F}_{*} P_{\Lambda_0} \,=\, P_{\Lambda_0}^{\mathcal{R}} \right\}} 2 \int [d\phi] \,P_{\Lambda_0}[\phi] \int d^d x \, d^d y \, \dot{C}_\Lambda^{-1}(x,y) \, \Big(\phi(x) - \mathcal{F}[\phi(x)]\Big)\Big(\phi(y) - \mathcal{F}[\phi(y)] \Big) \nonumber \\
& \qquad = \inf_{\left\{\mathcal{F} \,:\, \mathcal{F}_{*} P_{\Lambda_0} \,=\, P_{\Lambda_0}^{\mathcal{R}} \right\}} \mathsf{M}_{P_{\Lambda_0}}\![\mathcal{F}]
\end{align}
where $\mathcal{F}$ is a reparameterization from fields to fields and $\mathsf{M}_{P_{\Lambda_0}}$ is the analogue of the Monge functional in our setting.  Thus we can rewrite~\eqref{E:variational0} as
\begin{equation}
\mathcal{R}_{\tau} = \argmin_{\mathcal{R}} \left( \frac{1}{2 \tau}\,\inf_{\left\{\mathcal{F} \,:\, \mathcal{F}_{*} P_{\Lambda_0} \,=\, P_{\Lambda_0}^{\mathcal{R}} \right\}} \mathsf{M}_{P_{\Lambda_0}}\![\mathcal{F}] + S(P_{\Lambda_0}^{\mathcal{R}} \, \| \, Q_{\Lambda_0})\right)\,.
\end{equation}
Next we observe that that for a fixed $\mathcal{R}$, the infimum inside the $\argmin$ on the right-hand side will pick out an $\mathcal{F}$ such that $\mathcal{F}_{*} P_{\Lambda_0} \,=\, P_{\Lambda_0}^{\mathcal{R}}$.  As such, we can rewrite the above equation as
\begin{equation}
\mathcal{R}_{\tau} = \argmin_{\mathcal{R}} \inf_{\left\{\mathcal{F} \,:\, \mathcal{F}_{*} P_{\Lambda_0} \,=\, P_{\Lambda_0}^{\mathcal{R}} \right\}} \left( \frac{1}{2 \tau}\, \mathsf{M}_{P_{\Lambda_0}}\![\mathcal{F}] + S(\mathcal{F}_{*} P_{\Lambda_0} \, \| \, Q_{\Lambda_0})\right)\,.
\end{equation}
Since we are ultimately interested in having access to the RG-flowed distribution $P_{\Lambda_0}^{\mathcal{R}_{\tau}}$ and not necessarily $\mathcal{R}_\tau$ itself, the above equation suggests the following convenient reformulation: we have
\begin{equation}
P_{\Lambda_0 - \tau}[\phi] \approx \mathcal{F}_{\tau *} P_{\Lambda_0}[\phi]
\end{equation}
where $\mathcal{F}_{\tau}$ satisfies
\begin{equation}
\label{E:Feq0}
\boxed{\mathcal{F}_\tau = \argmin_{\mathcal{F}} \left( \frac{1}{2 \tau}\, \mathsf{M}_{P_{\Lambda_0}}\![\mathcal{F}] + S(\mathcal{F}_{*} P_{\Lambda_0} \, \| \, Q_{\Lambda_0})\right)}
\end{equation}
Moreover this should become exact as $\tau \to 0$.  This equation is more convenient than~\eqref{E:variational0} since it only has a single minimization, i.e.~we have successfully accommodated for the reparameterization minimization and the Wasserstein-2 minimization in one fell swoop.

Our reformulation of~\eqref{E:variational0} into~\eqref{E:Feq0} can similarly be applied to~\eqref{E:variational1}.  In particular, let $\mathcal{F}_\tau, \mathcal{F}_{2\tau}, \mathcal{F}_{3\tau},...$ be a sequence of reparameterizations and define
\begin{equation}
\widetilde{\mathcal{F}}_{n \tau} := \mathcal{F}_\tau \circ \mathcal{F}_{2\tau} \circ \cdots \circ \mathcal{F}_{n \tau}\,.
\end{equation}
Note the ordering of the composition relative to~\eqref{E:Rtilde1}, since we can think of the $\mathcal{F}$'s as acting inversely as the $\mathcal{R}$'s.  Then we have
\begin{align}
P_{\Lambda_0 - n \tau}[\phi] &\approx \widetilde{\mathcal{F}}_{n \tau *} P_{\Lambda_0}[\phi] \nonumber \\
&= (\mathcal{F}_{n\tau *} \circ \cdots \circ \mathcal{F}_{2\tau *} \circ \mathcal{F}_{\tau *}) P_{\Lambda_0}[\phi]
\end{align}
where the $\mathcal{F}_{n \tau}$'s satisfy
\begin{equation}
\label{E:Feq1}
\boxed{\mathcal{F}_{(n+1)\tau} = \argmin_{\mathcal{F}} \left( \frac{1}{2 \tau}\, \mathsf{M}_{\widetilde{\mathcal{F}}_{n \tau *} P_{\Lambda_0}}\![\mathcal{F}] + S((\mathcal{F}_{*} \circ \widetilde{\mathcal{F}}_{n \tau *}) P_{\Lambda_0} \, \| \, Q_{\Lambda_0 - n \tau})\right)}
\end{equation}
This is the desired generalization of~\eqref{E:variational1} which only has a single optimization.

In the next subsection, we will explore strategies for solving~\eqref{E:Feq1} via numerical optimization.  For the moment, let us unpack~\eqref{E:Feq1} slightly, and write it in a more convenient form.  First, by iteratively changing integration variables using the diffeomorphisms $\mathcal{F}_{n\tau}$, we can rewrite $\mathsf{M}_{\widetilde{\mathcal{F}}_{n \tau *} P_{\Lambda_0}}\![\mathcal{F}]$ as
\begin{align}
\label{E:Mexpect1}
&\mathsf{M}_{\widetilde{\mathcal{F}}_{n \tau *} P_{\Lambda_0}}\![\mathcal{F}] \nonumber \\
&\,\, = 2\int[d\phi]\,P_{\Lambda_0}[\phi] \int d^d x \, d^d y \, \dot{C}_{\Lambda}^{-1}(x,y) \Big(\widetilde{\mathcal{F}}_{n \tau}[\phi(x)] - (\mathcal{F} \circ \widetilde{\mathcal{F}}_{n \tau})[\phi(x)]\Big)\Big(\widetilde{\mathcal{F}}_{n \tau}[\phi(y)] - (\mathcal{F} \circ \widetilde{\mathcal{F}}_{n \tau})[\phi(y)] \Big) \nonumber \\
&\,\, = \E_{P_{\Lambda_0}}\!\left[2\int d^d x \, d^d y \, \dot{C}_{\Lambda}^{-1}(x,y) \Big(\widetilde{\mathcal{F}}_{n \tau}[\phi(x)] - (\mathcal{F} \circ \widetilde{\mathcal{F}}_{n \tau})[\phi(x)]\Big)\Big(\widetilde{\mathcal{F}}_{n \tau}[\phi(y)] - (\mathcal{F} \circ \widetilde{\mathcal{F}}_{n \tau})[\phi(y)] \Big)\right]\,.
\end{align}
In a similar fashion, we can write
\begin{align}
\label{E:Srelexpect1}
S((\mathcal{F}_{*} \circ \widetilde{\mathcal{F}}_{n \tau *}) P_{\Lambda_0} \, \| \, Q_{\Lambda_0 - n \tau}) &= \int [d\phi_0]\,P_{\Lambda_0} \log\left(P_{\Lambda_0}/Q_{\Lambda_0 - n\tau}^{\widetilde{\mathcal{F}}_{n \tau} \circ \mathcal{F}}\right) \nonumber \\
&= \E_{P_{\Lambda_0}}\!\left[\log\left(P_{\Lambda_0}/Q_{\Lambda_0 - n\tau}^{\widetilde{\mathcal{F}}_{n \tau} \circ \mathcal{F}}\right)\right]\,.
\end{align}
This change of variables follows from iteratively utilizing the infinite-dimensional analog of the reparametrization-invariance of the relative entropy, which in finite dimensions is the statement that $S(f_*p\,\|\,q) = S(p,\|\, f^{-1}_*q)$ for probability distributions $p,q$ and a diffeomorphism $f$.

Combining~\eqref{E:Mexpect1} and~\eqref{E:Srelexpect1}, we can write~\eqref{E:Feq1} in the form
\begin{equation}
\label{E:lossfunction1}
\boxed{\mathcal{F}_{(n+1)\tau} = \argmin_{\mathcal{F}}\, \E_{P_{\Lambda_0}}\textsf{Loss}\!\left[\mathcal{F}\,,\,\widetilde{\mathcal{F}}_{n \tau}\,,\,P_{\Lambda_0}\,,\,Q_{\Lambda_0 - n \tau}\right]}
\end{equation}
where $\textsf{Loss}\!\left[\mathcal{F}\,,\,\widetilde{\mathcal{F}}_{n \tau}\,,\,P_{\Lambda_0}\,,\,Q_{\Lambda_0 - n \tau}\right]$ is the function to be minimized.\footnote{Calling this the `loss' function is standard in the computer science literature.}

\subsection{Numerical applications of variational formulae}
\label{Sec:NumericalMethods}

The variational characterization of RG flows discussed above suggests new and interesting numerical methods for (approximately) computing such flows.  In particular, suppose we have sample access to $P_{\Lambda_0}[\phi]$.  For the purposes of this section, we will take our fields to be lattice-discretized on a finite volume domain; then we can sample from $P_{\Lambda_0}[\phi]$ by employing standard Monte Carlo methods.  Equation~\eqref{E:lossfunction1} tells us that such sampling access is sufficient in principle to solve for $\mathcal{F}_{\tau}, \mathcal{F}_{2\tau}, ..., \mathcal{F}_{n \tau}$.  We will return shortly to the problem of how the requisite minimizations can be implemented in practice.  For the moment, let us say we have $\mathcal{F}_{\tau}, \mathcal{F}_{2\tau}, ..., \mathcal{F}_{n \tau}$ at hand, in which case we would like to be able to sample from the RG-flowed distribution $P_{\Lambda_0 - n \tau}[\phi] \approx (\mathcal{F}_{n\tau *} \circ \cdots \circ \mathcal{F}_{2\tau *} \circ \mathcal{F}_{\tau *}) P_{\Lambda_0}[\phi]$.  How can we sample from such a distribution?  Fortunately, sampling is readily compatible with the pushforward operation, as the following algorithm demonstrates:
\begin{algorithm}[h!]
   \caption{Sampling from $(\mathcal{F}_{n\tau *} \circ \cdots \circ \mathcal{F}_{2\tau *} \circ \mathcal{F}_{\tau *}) P_{\Lambda_0}$}
\begin{algorithmic}
    \STATE {\bfseries Input:} Reparameterizations $\mathcal{F}_{\tau}, \mathcal{F}_{2\tau},...,\mathcal{F}_{n\tau}$, sample access to $P_{\Lambda_0}$
    \STATE {\bfseries Output:} A sample $\widehat{\phi}$ from $(\mathcal{F}_{n\tau *} \circ \cdots \circ \mathcal{F}_{2\tau *} \circ \mathcal{F}_{\tau *}) P_{\Lambda_0}$
    \vspace{0.25em}
    \STATE Sample $\phi \leftarrow P_{\Lambda_0}$
    \STATE Compute $\widehat{\phi} = (\mathcal{F}_{\tau} \circ \mathcal{F}_{2\tau} \circ \cdots \circ \mathcal{F}_{n\tau})[\phi]$
    \STATE \textbf{return} $\hat{\phi}$
\end{algorithmic}
\end{algorithm}

Now we turn to the more interesting problem of numerically solving~\eqref{E:lossfunction1} for the reparameterizations $\mathcal{F}_{\tau}, \mathcal{F}_{2\tau}, ...$.  A natural way to proceed is to let our reparameterizations $\mathcal{F}_{n \tau}$ have a particular form that only depends on a finite-dimensional vector of real parameters $\theta_n$\,; we write this dependence as $\mathcal{F}_{n \tau} = \mathcal{F}_{\theta_{n}}$.  Moreover, we define
\begin{equation}
\widetilde{\mathcal{F}}_{\theta_1,...,\theta_n} := \mathcal{F}_{\theta_1} \circ \mathcal{F}_{\theta_2} \circ \cdots \circ \mathcal{F}_{\theta_n}\,.
\end{equation}
In this setting,~\eqref{E:lossfunction1} becomes
\begin{equation}
\label{E:thetalossfunction1}
\theta_{n+1} = \argmin_{\theta} \E_{P_{\Lambda_0}}\textsf{Loss}\!\left[\mathcal{F}_{\theta}\,,\,\widetilde{\mathcal{F}}_{\theta_1,...,\theta_n}\,,\,P_{\Lambda_0}\,,\,Q_{\Lambda_0 - n \tau}\right]
\end{equation}
If $\mathcal{F}_{\theta}$ is a differentiable function of $\theta$, then we can bring to bear techniques from machine learning to perform the optimization in~\eqref{E:thetalossfunction1}.

First we ask: what is a good family of functionals $\mathcal{F}_{\theta}$ to choose?  There are certain requirements that the family should have.  For instance, if we work with a translationally invariant field theory, then we would like for
\begin{equation}
\mathcal{F}_{\theta}[\phi(y+a)](x) = \mathcal{F}_{\theta}[\phi(y)](x+a)
\end{equation}
for all $\theta$.  This is an equivariance condition on $\mathcal{F}_{\theta}$ with respect to translations.  Moreover, in spatially local RG schemes such as those we studied in the context of Polchinski's equation and the Wegner-Morris flow equation, we would also like $\mathcal{F}_{\theta}[\phi(y)](x)$ to only depend strongly on values of $\phi(y)$ near $y \approx x$.

These properties imply that a good ansatz for $\mathcal{F}_\theta$ is to let it be a convolutional neural network~\cite{Goodfellow-et-al-2016} with weights $\theta$. Indeed, in the context of numerical algorithms, convolutional neural networks are known~\cite{imagenet} to give good ansatzes for general, translationally invariant functionals of fields such that the functional is spatially local in the sense we discussed above.  For concreteness, we remind the reader that if $\phi$ is, for example, a lattice discretization $\phi(i,j)$ of a two-dimensional field, then a convolutional neural network of depth $D$ would render $\mathcal{F}_\theta[\phi]$ as having the form
\begin{align}
\label{eq:neural-network}
    \mathcal{F}^0_\theta[\phi](i, j) &= \phi(i, j) \\
    \mathcal{F}^{\ell}_\theta[\phi](i, j) &= \sigma \!\left( \sum_{m, n= -k}^k A^\ell_{m, n} \mathcal{F}^{\ell-1}[\phi](i+m, j+n) \right)\\
 \mathcal{F}^D_\theta[\phi] &= \mathcal{F}_\theta[\phi]
\end{align}
for $\ell = 1,...,D$ where the $A^\ell_{mn}$ are $2k \times 2k$ matrices representing discrete convolutional kernels, while $\sigma(x)$ is a nonlinear function such as $\sigma(x) = \frac{1}{1+e^{-x}}$. The parameters $\theta$ of the neural network are the entries of the matrix kernels $A^1, ..., A^D$. We note also that $k$ can depend on the ``layer'' $\ell$, and that often one inserts intermediate ``max-pool'' or ``average-pool'' layers that down-sample the intermediate fields $\mathcal{F}_{\theta}^\ell[\phi]$. For example, in average-pooling,  one replaces $\mathcal{F}_{\theta}^\ell[\phi](i, j)$ by a new function
\begin{equation}
    \widetilde{\mathcal{F}^\ell_\theta[\phi]}(i', j') = \frac{1}{|B(i', j')|} \sum_{(i, j) \in B(i', j')} \mathcal{F}^\ell_\lambda[\phi](i,j)
\end{equation}
where $(i', j')$ runs over a lattice with fewer sites, and $B(i', j')$ is a subset of the indices of the $(i, j)$ lattice, exactly as in block spin renormalization.

With a particular neural network architecture for $\mathcal{F}_{\theta}$ in mind, we examine how to solve~\eqref{E:thetalossfunction1}. Observe that the quantity to be minimized in that equation is an expectation value over $P_{\Lambda_0}$.  Therefore, the gradient of this quantity with respect to the variational parameter $\theta$ is also an explicit expectation value over $P_{\Lambda_0}$. Accordingly, if we have sampling access to $P_{\Lambda_0}$\,, then as is standard in stochastic gradient descent, we can approximate $\E_{P_{\Lambda_0}}\!\!\nabla_\theta\, \textsf{Loss}$ via a Monte Carlo approximation, replacing the expectation value with an evaluation of the gradient on a single sample from $P_{\Lambda_0}$\,. The nontrivial fact that the quantity to be optimized is an expectation value over $P_{\Lambda_0}$ (which holds due to the reparametrizations in \eqref{E:Mexpect1}, \eqref{E:Srelexpect1}) is needed to even conceive of a plausible numerical algorithm in this setting, since general integrals over the space of fields are completely intractable. Indeed, the problem of turning this variational formulation of RG flow into a tractable numerical method is rather interesting, and must involve the use of several approximation techniques from neural networks beyond the most basic formulation given above.\footnote{In particular, the efficient discretization of the functional determinant $\left|\frac{\delta \mathcal{F}_\theta[\phi]}{\delta \phi}\right|$ and its $\theta$-derivatives is tricky; the natural lattice analog of the functional determinant is  $|\det \nabla_{\phi_{ij}} \mathcal{F}_\theta|$, which does not make sense for RG flows which decrease the number of effective lattice sites. The simplest solution is to let $\mathcal{F}_\theta$ be a convolution that does not decrease the number of lattice sites, and to utilize the ResNet technique~\cite{resnet}.  This is a standard technique in neural networks which involves the replacement $\mathcal{F}_\theta \to \text{Id} + \mathcal{F}_\theta$, which for small $\theta$ forces the Jacobian to be an invertible matrix, making the log-determinant nondegenerate. This neural network architecture is also natural since $\mathcal{F}_\theta$ should approximate a small renormalization group transformation, which should be thought of as a perturbation of the identity transformation.  Finally, the log determinant and its gradient are challenging quantities to compute efficiently when the dimension is large; however, there are efficient tricks to analytically compute log determinants of convolutions using circulant matrices~\cite{invertible-convolutional-flows}, which can also be utilized to force $\mathcal{F}_\theta$ to be a diffeomorphism.}  We leave to future work a comprehensive exploration of neural network numerical methods based on the variational formulation of RG flow.

In more detail, the stochastic gradient descent algorithm for computing $\theta_{n+1}$ is:
\begin{algorithm}[h!]
   \caption{Stochastic gradient descent for computing $\theta_{n+1}$}
\begin{algorithmic}

    \STATE {\bfseries Input:} Sampling access to $P_{\Lambda_0}$, loss function $\textsf{Loss}\!\left[\mathcal{F}_{\theta}\,,\,\widetilde{\mathcal{F}}_{\theta_1,...,\theta_n}\,,\,P_{\Lambda_0}\,,\,Q_{\Lambda_0 - n \tau}\right] =: \textsf{Loss}[\theta, \phi]$ 
    \STATE \qquad \quad \,\, parameterized as a neural network in a differentiable parameter $\theta$, initial $\theta$ value $\theta_{\text{init}}$,
    \STATE \qquad \quad \,\, maximum step number $\tau_{\text{max}}$, rate parameter $r$
    \STATE {\bfseries Output:} Approximation to $\theta_{n+1}$
    \vspace{0.25em}
    \STATE Initialize $\theta = \theta_{\text{init}}$
    \FOR{$\tau = 1, \dots, \tau_\mathrm{max}$}
      \STATE Sample $\phi \leftarrow P_{\Lambda_0}$
    \STATE Compute $\nabla_\theta\, \textsf{Loss}[\theta,\phi]$ via backpropagation
    \STATE Replace $\theta \to \theta - r \,\nabla_\theta\, \textsf{Loss}[\theta,\phi]$
    \ENDFOR
    \STATE \textbf{return} $\theta$
\end{algorithmic}
\end{algorithm}

\noindent We note that the relative entropy contained in $\textsf{Loss}[\theta,\phi]$ is in fact finite due to our (hard) lattice cutoff and finite volume domain.  This is in contrast to our continuum analysis of the relative entropy in Section~\ref{Sec:Monotones}, in which we had to contend with divergences and associated subtleties with orders of limits.

There is a wealth of ideas that have circulated about connections between the renormalization group and the hierarchical structure of convolutional neural networks~\cite{dlrg1, dlrg2, dlrg3, dlrg4, dlrg5, dlrg6, dlrg7}. These connections have at times informed the theory and methodology around neural network training~\cite{rgdlapp1, rgdlapp2}. However, these considerations were largely heuristic, and did not connect convolutional neural networks with any explicit renormalization group flow for any particular theory. The formulation provided in this section seems to be the first precise connection between an optimization problem based on convolutional neural networks and an explicit instantiation of the renormalization group in the standard setting of field theory.

\section{Discussion}
\label{Sec:Discussion}

In this paper we have provided a new approach to the exact renormalization group using the tools of optimal transport theory.  In so doing, we defined new, non-perturbative RG monotones, developed a novel variational formula for RG flows, and suggested new numerical algorithms.

Going forward, it would be interesting to apply the techniques in this paper to a richer class of field theories, such as gauge theories (see e.g.~\cite{Reuter:1993kw, Pawlowski:2005xe, Gies:2006wv}).  Moreover, it would be desirable to compute more examples of our RG monotone in any setting.  In the realm of scalar field theories, a natural target would be to consider flows in the neighborhood of the Wilson-Fisher fixed point.

Since we have defined RG monotones for a large class of RG flows, it seems possible that for a judicious choice of RG flow  (for instance, a judicious choice of the seed action $\hat{S}_\Lambda[\phi]$ in the Wegner-Morris formulation) one could use the positivity of our monotone to constrain the signs of couplings in effective field theory.  

It would be very interesting to empirically investigate numerical methods based on our proposal in Section~\ref{Sec:NumericalMethods}.  The initial neural-network-based proposal that we describe connects nicely with recent advances in neural network approaches to Wasserstein gradient flows~\cite{wasserstein_gradient_neural_1, wasserstein_gradient_neural_2}.  The use of specialized neural network algorithms~\cite{wasserstein_gradient_neural_3}, possibly coupled with connections to fast approximations of the Wasserstein distance~\cite{sinkhorn}, may allow one to robustly approximate RG flows with relatively few discrete steps. It is an important practical problem to make the initial numerical method proposed above significantly more numerically efficient using the wealth of ideas in the machine learning literature on generative models and normalizing flows; our proposal is only a first step towards a practical numerical method.  

Since RG allows one to determine $P_\Lambda$ for different scales $\Lambda$, the variational formulation of RG may allow for improved sampling algorithms via connections to recent advances in neural network generative models~\cite{neural_diffusion}, which involve denoising procedures that are heuristically related to the inversion of the renormalization group flow.  A related generalization would be to develop optimal transport algorithms for continuous MERA (cMERA) tensor networks~\cite{haegeman2013entanglement, cotler2019renormalization, cotler2019entanglement, Hu:2018hyd}, by leveraging and generalizing the known connection between cMERA and ERG~\cite{Fliss:2016ifp}.  A suggestive possibility is to implement a backwards gradient flow to go from an IR cMERA ansatz to a UV state. In a similar vein, perhaps one could adapt the optimal transport technology to study the RG flow of quantum states using techniques from (see e.g.~\cite{carlen2014analog, carlen2017gradient}).

More broadly, it seems that many more tools from optimal transport, possibly combined with information theory, can be brought to bear on the subject of ERG via our present formulation.  For instance, it appears likely that our formulation of RG flow in this manuscript could be synthesized with the approaches of~\cite{Beny:2012qh, Beny:2014sna, Balasubramanian:2014bfa, lashkari2019entanglement, Furuya:2020tzv, Fowler:2021oje, Koenigstein:2021rxj, Erdmenger:2021sot} from the physics community.  Since the optimal transport community has enormous analytical and numerical traction in the PDE setting, it would be valuable to adapt these insights to the functional generalizations appropriate for ERG flows.

\subsection*{Acknowledgements}
We thank Kristan Jensen, Igor Klebanov, Nima Lashkari, Tim Morris, Yair Shenfeld, and Andrew Strominger for valuable discussions.  We give a special thanks to Arthur Kosmala for identifying and correcting an error in our definition of the Wasserstein distance in the quantum field theory setting.
JC is supported by a Junior Fellowship from the Harvard Society of Fellows, the Black Hole Initiative, as well as in part by the Department of Energy under grant {DE}-{SC0007870}. SR is supported by the Simons Foundation Collaboration grant ``Homological Mirror Symmetry and Applications'' (Award \# 385573). 

\appendix

\section{Comments on the infinitesimal form of the $\mathcal{W}_2$ metric}
\label{App:heuristic}

This Appendix summarizes a derivation of the infinitesimal form of the Wasserstein metric from its finite-distance definition, clarifying the inversion of the Riemannian metric that occurs when passing from the finite-distance form to the infinitesimal form. An infinite-dimensional analog of this computation leads to the functional Wasserstein metric of \eqref{E:functionaltotalW2}.

For the purposes of this Appendix, it is useful to consider a modified version of the heat equation
\begin{align}
\label{E:modifiedheat1}
\frac{\partial p}{\partial t} = A^{ij} (\partial_i \partial_j p)\,,
\end{align}
where $A^{ij}$ is positive semi-definite as a matrix.  Throughout this Appendix, we will use Einstein index notation so that $A^{ij}$ and $A_{ij}$ are inverses, i.e.~$A_{ij}A^{jk} =\delta_i^k$.

Let $\text{dens}(M)$ be the space of probability distributions on $M = \mathbb{R}^d$ so that the tangent space is $T_p \text{dens}(M) = \{\bar{\eta} \in C^\infty(\mathbb{R}^d)\, : \, \int dx \, \bar{\eta} = 0\}$ for any $p \in \text{dens}(M)$.  For any tangent vector $\eta$ in $T_p \,\text{dens}(M)$ we have an associated $\bar{\eta}$ obtained by solving
\begin{equation}
\label{eq:dualizing-equation-2}
    A^{ij} \,\partial_i( p \,\partial_j\bar{\eta})= \eta\,.
\end{equation}
This solution $\bar{\eta}$ is unique up to an additive constant; this induces the identification $\eta \leftrightarrow \bar{\eta}$ via an isomorphism
\begin{equation}
T_p\,\text{dens}(M) \simeq \overline{T_p\,\text{dens}(M)} := \{ \bar{\eta} \in C^\infty(M)\}/\{\text{constants}\}\,. 
\end{equation}
With this isomorphism in mind, we can write down the Riemannian metric 
\begin{equation}
\label{eq:infinitesimal-wasserstein-metric-2}
\langle \eta_1, \eta_2 \rangle_{\mathcal{W}_2} := \int dx\,p\,A^{ij}\,\partial_i\bar{\eta}_1\,\partial_j\bar{\eta}_2 = -\int dx\, \eta_1 \bar{\eta}_2 = -\int dx\, \bar{\eta}_1 \eta_2 \,.
\end{equation}
The last two equalities are obtained via integration by parts.  Using this Riemannian metric we have that the modified heat equation in~\eqref{E:modifiedheat1} can be written as
\begin{equation}
\frac{\partial p}{\partial t} = \nabla_{\mathcal{W}_2} S[p]\,,
\end{equation}
since here $\nabla_{\mathcal{W}_2} S[p] = A^{ij}(\partial_i \partial_j p)$.  A rigorous argument given in~\cite[Lemma 4.3]{otto2005eulerian} establishes that~\eqref{eq:infinitesimal-wasserstein-metric-2} is the infinitesimal version of the Wasserstein-2 metric
\begin{equation*}
\mathcal{W}_2(p_0,p_1) := \left(\inf_{\pi \in \Gamma(p_0, p_1)} \int dx \, dy \,\pi(x,y) \, A_{ij}\,(x^i-y^i)(x^j-y^j) \right)^{1/2}\,,
\end{equation*}
and here we will explain some heuristics for key parts of the proof.

Recall that in Riemannian geometry, given a path $x(u)$ with $u\in[0,1]$ in a Riemannian manifold, the length of that path is given by
\begin{equation}
L[x(u)] = \int_0^1 du \, \sqrt{A_{ij}\,\partial_u x^i(u) \,\partial_u x^j(u)}\,.
\end{equation}
The minimizers of the length functional $L[x(u)]$ with fixed boundary conditions at $x(0)$ and $x(1)$ are geodesics.  However, these minimizers are always non-unique because the length functional is invariant under reparameterizations.  This high degree of non-uniqueness can be avoided by instead considering the energy functional 
\begin{equation}
 E[x(u)] = \frac{1}{2} \int_0^1 du\, A_{ij}\,\partial_u x^i(u) \,\partial_u x^j(u) \,.
\end{equation}
Its minimizers with fixed boundary conditions at $x(0)$ and $x(1)$ are exactly geodesics with constant speed.  (In essence, the energy functional picks out a preferred `reparameterization' of $x(u)$).  Now using the Cauchy-Schwarz inequality, we have 
\begin{equation}
\label{E:LEineq1}
    L[x(u)]^2 \leq 2 E[x(u)] 
\end{equation}
with equality exactly when $|x'(u)|$ is constant in time, i.e.~$x(u)$ is parameterized so it has constant speed.  This implies that
\begin{equation}
\label{E:LvsE1}
\inf_{\{x(u) \, : \, x(0) \,=\, a\,,\,x(u) \,=\, b\}} L[x(u)]^2 = \inf_{\{x(u) \, : \, x(0) \,=\, a\,,\,x(1) \,=\, b\}} 2 E[x(u)]\,,
\end{equation}
namely that $L[x(u)]^2$ and $2 E[x(u)]$ have the same minimizing values.

We will apply the above insights to study the Riemannian metric~\eqref{eq:infinitesimal-wasserstein-metric-2} on $T_p\,\text{dens}(M)$.  Suppose we have a 1-parameter family of probability distributions $p(u)$ for $u\in[0,1]$ where we take $p(0) = p_0$ and $p(1) = p_1$.  We emphasize that the $u$ appearing in $p(u)$ (which we will also write as $p(x,u)$) is \textit{different} from the time coordinate $t$ appearing in~\eqref{E:modifiedheat1}.  The $t$ there corresponds to time evolution, whereas the $u$ here parametrizes a geodesic flow in the space of probability distributions.  With this in mind, we define $\phi(u)$ as a 1-parameter family of solutions to the equations
\begin{equation}
\label{E:family1}
    \frac{\partial}{\partial u} \,p(u) = A^{ij}\,\partial_i(p(u) \,\partial_j\phi)\,.
\end{equation}
The energy of the path $p(u)$ with respect to the Riemannian metric~\eqref{eq:infinitesimal-wasserstein-metric-2} is given by 
 \begin{equation}
    E_{\mathcal{W}_2}[p(u)] :=  \frac{1}{2}\int_0^1 du\int dx \,p(u)\,A^{ij} \,\partial_i \phi \, \partial_j \phi\,.
 \end{equation}
Writing $V^i(x,u) = -A^{ij}\, \partial_j\phi(x,u)$, we can define a flow on $\text{dens}(M)$, namely
 \begin{equation}
 \Phi_{u\,*} : \text{dens}(M) \longrightarrow \text{dens}(M)\,, \qquad u \in [0,1]\,,
 \end{equation}
 via the differential equation
 \begin{equation}
 \label{E:Phidef1}
\frac{\partial}{\partial u}\,\Phi_u = V(\Phi_u\,, u)\,,\qquad \Phi_0 = \text{Id}\,.
 \end{equation}
Let us check that $p(u) = \Phi_{u \,*} \,p_0$\,.  It suffices to show that
\begin{equation}
\label{E:toestablish1}
p(x, u + du) = p(\Phi_{-du}(x),u) \, |\det \, \partial \Phi_{-du} |\,.
\end{equation}
We first note that
\begin{align}
\Phi_{-du}^i(x) &= x^i - du\, V^i \nonumber \\
&= x^i + du\,A^{ik}\,\partial_k \phi(x,u)
\end{align}
and also
\begin{align}
\partial_j\Phi_{-du}^i(x) &= \delta_j^i + du\, A^{ik}\,\partial_j\partial_k \phi(x,u)\,.
\end{align}
It follows that
\begin{align}
p(\Phi_{-du}^i(x),u)  &= p(x^i + du\,A^{ij}\,\partial_j \phi(x,u), u) \nonumber \\
&= p(x,u) + dt\, \partial_i p(x,u) \, A^{ij}\,\partial_j \phi(x,u)
\end{align}
and accordingly
\begin{align}
|\det \, \partial \Phi_{-du} | &= 1 + du \, A^{ij}\, \partial_i \partial_j \phi(x,u)\,.
\end{align}
Altogether we have
\begin{align}
p(\Phi_{-du}(x),u) \, |\det \, \partial \Phi_{-du} | &= p(x,u) + du\,A^{ij}\,\partial_i\!\left(p(x,u)\,\partial_j \phi(x,u)\right) \nonumber \\
&=p(x,u) + du\, \frac{\partial}{\partial u}\,p(x,u)
\end{align}
where we have used~\eqref{E:family1} in going from the first line to the second line.  This establishes~\eqref{E:toestablish1}.

Having checked that $p(u) = \Phi_{u \,*} \,p_0$, we have the standard inequalities
\begin{align}
\sqrt{A_{ij}\,(x -\Phi_1(x))^i\,(x -\Phi_1(x))^j} &= \sqrt{A_{ij}\,(\Phi_0(x) -\Phi_1(x))^i\,(\Phi_0(x) -\Phi_1(x))^j} \nonumber \\
&\leq \int_0^1 du\,\sqrt{A_{ij}\,\left(\frac{\partial}{\partial u} \,\Phi_u(x)\right)^i \,\left(\frac{\partial}{\partial u} \,\Phi_u(x)\right)^j}\,.
\end{align}
In words, this equality holds because the metric distance between the endpoints of a curve is upper bounded by the length of that curve.  Combining this with the inequality~\eqref{E:LEineq1}, we obtain
\begin{equation}
\label{E:usefulineq1}
A_{ij}\,(x -\Phi_1(x))^i\,(x -\Phi_1(x))^j \leq \int_0^1 du\,A_{ij}\,\left(\frac{\partial}{\partial u} \,\Phi_u(x)\right)^i \,\left(\frac{\partial}{\partial u} \,\Phi_u(x)\right)^j\,.
\end{equation}
This inequality~\eqref{E:usefulineq1} will be useful to us below.

Let us show that $\frac{1}{2}\,\mathcal{W}_2(p_0,p_1)^2 \leq E_{\mathcal{W}_2}[p(u)]$.  Using the definition of $\mathcal{W}_2(p_0, p_1)$, we have
\begin{align}
\label{E:KtoM1}
\frac{1}{2}\,\mathcal{W}_2(p_0,p_1)^2 &= \frac{1}{2}\inf_{\pi \in \Gamma(p_0, p_1)} \int dx \, dy \,\pi(x,y) \, A_{ij}\,(x^i - y^i)(x^j - y^j) \nonumber \\
&\leq \frac{1}{2} \int dx \, dy \,\,p_0(x) \delta(y - \Phi_1(x)) \, A_{ij}\,(x^i - y^i)(x^j - y^j)  \nonumber \\
&= \frac{1}{2} \int dx \, p_0(x)\,A_{ij}\,(x - \Phi_1(x))^i \,(x - \Phi_1(x))^j \,.
\end{align}
This inequality comes from making the particular choice of $\pi(x,y) = p_0(x) \delta(y - \Phi_1(x))$, which may not be the minimizing choice of $\pi(x,y)$.  Next, we use~\eqref{E:usefulineq1} to upper bound the last line of~\eqref{E:KtoM1} as
\begin{align}
\label{E:upperbound2}
\frac{1}{2} \int dx \, p_0\,A_{ij}\,(x -\Phi_1(x))^i\,(x -\Phi_1(x))^j \leq \frac{1}{2}\int dx \int_0^1 du \, p_0\,A_{ij}\,\left(\frac{\partial}{\partial u} \,\Phi_u(x)\right)^i \,\left(\frac{\partial}{\partial u} \,\Phi_u(x)\right)^j \,.
\end{align}
Using the definition of~\eqref{E:Phidef1} and the fact that $V(x,u) = -A^{ij}\,\partial_j\phi(x,u)$, the right-hand side of the above inequality equals
\begin{align}
\label{E:usefulequalities}
\frac{1}{2}\int dx \int_0^1 du \, p_0\,A^{ij} \, \partial_i \phi(\Phi_u(x),u) \, \partial_j \phi(\Phi_u(x),u)  &= \frac{1}{2} \int dx \int_0^1 du \, \Phi_{u\,*}p_0\,A^{ij} \, \partial_i \phi(x,u) \, \partial_j \phi(x,u) \nonumber \\
&= \frac{1}{2} \int dx \int_0^1 du \, p(u)\,A^{ij} \, \partial_i \phi(x,u) \, \partial_j \phi(x,u) \nonumber \\
&= E_{\mathcal{W}_2}[p(u)]\,.
\end{align}
Combining~\eqref{E:KtoM1},~\eqref{E:upperbound2}, and~\eqref{E:usefulequalities} we obtain the desired inequality
\begin{equation}
\frac{1}{2}\, \mathcal{W}_2(p_0, p_1)^2 \leq E_{\mathcal{W}_2}[p(u)]\,.    
\end{equation}
One works harder to show that, in fact,
\begin{equation}
\frac{1}{2}\, \mathcal{W}_2(p_0, p_1)^2 = \inf_{\{p(u) \, : \, p(0) \,=\, p_0\,,\,p(1) \,=\, p_1\}} E_{\mathcal{W}_2}[p(u)]\,.
\end{equation}
Defining the length
\begin{equation}
    L_{\mathcal{W}_2}[p(u)] = \int_0^1 du \int dx \sqrt{ p \,A^{ij}\,\partial_i \phi \, \partial_j \phi }
\end{equation}
and applying a similar logic as that which led to~\eqref{E:LvsE1}, we find
\begin{equation}
\inf_{\{p(u) \, : \, p(0) \,=\, p_0\,,\,p(1) \,=\, p_1\}} L_{\mathcal{W}_2}[p(u)]^2 := \inf_{\{p(u) \, : \, p(0) \,=\, p_0\,,\,p(1) \,=\, p_1\}} 2 E_{\mathcal{W}_2}[p(u)]
\end{equation}
and thus
\begin{equation}
\mathcal{W}_2(p_0, p_1) = \inf_{\{p(u) \, : \, p(0) \,=\, p_0\,,\,p(1) \,=\, p_1\}} L_{\mathcal{W}_2}[p(u)]\,.
\end{equation}
This establishes that the metric~\eqref{eq:infinitesimal-wasserstein-metric-2} is in fact the infinitesimal form of the Wasserstein-2 metric.

\section{Convenient cutoff functions}
\label{App:convenient}
In order for the derivative $- \Lambda \frac{d}{d\Lambda} M_\Lambda$ of our RG monotone quantity to be perturbatively or nonperturbatively finite, we saw in Section 5.1 that we must have 
\begin{equation}
\label{E:integral-condition}
\int d^dp \,\frac{\partial \log K_\Lambda(p^2)}{\partial \Lambda} < \infty\,.
\end{equation}
This finiteness property does not hold for all cutoff functions $K_\Lambda(p^2)$; for instance, it fails to hold for
\begin{equation}
\label{E:standard-cutoff-function}
K(p^2/\Lambda^2) = \frac{1 + e^{-a}}{1 + e^{a\left((p/\Lambda)^2 - 1\right)}}
\end{equation}
where $a$ is a constant, usually taken to be much greater than one.

The reason that cutoff functions like~\eqref{E:standard-cutoff-function} lead to divergent $\int d^d p \, \frac{\partial \log K_\Lambda(p^2)}{\partial \Lambda}$ is that for such cutoff functions, as one varies $\Lambda$, arbitrarily high scales are suppressed by a multiplicative factor which does not decay appreciably with $p^2$. In particular, for any $\Lambda$, the logarithmic derivative of the $K_\Lambda$ in~\eqref{E:standard-cutoff-function} is greater than $1/|p|^d$ for all sufficiently large $p^2$, and so the integral $\int d^d p \, \frac{\partial \log K_\Lambda(p^2)}{\partial \Lambda}$ diverges. The condition~\eqref{E:integral-condition} should thus be interpreted as requiring that the factor by which one suppresses large frequencies as one varies $\Lambda$ infinitesimally should decay rapidly with $p^2$.  This condition is readily satisfied in an infinite family of cutoff functions.

In order to construct this desirable class of cutoff functions, it is instructive to first understand what other properties a cutoff function should have.  Firstly, we would like $K_\Lambda(p^2)$ to decay faster than any polynomial as a function of $p^2$ for $p^2 \gtrsim \Lambda^2$ in order for perturbation theory of our field theory to be sensible when the cutoff scale is $\Lambda$.  Secondly, $K_\Lambda(p^2)$ should be very close to one for $p^2 \leq \Lambda^2 - \varepsilon$ where $\varepsilon > 0$ is a small constant, and $K_\Lambda(p^2)$ should drop off to near zero for $p^2 \geq \Lambda^2 + \varepsilon$.  For simplicity, we impose the following requirements:
\begin{equation}
\label{eq:cutoff-requirements}
    \begin{minipage}{0.9\textwidth}
    \begin{enumerate}
    \item $K_\Lambda(p^2) = 1$ for $p^2 \leq \Lambda^2$\,,
    \item $K_\Lambda(p^2)$ is monotonically decreasing for $p^2 \geq \Lambda^2$\,,
    \item $K_\Lambda(p^2) \leq \varepsilon$ for $p^2 \geq \Lambda^2 + \varepsilon$\,,
    \item $\int d^dp \,\frac{\partial \log K_\Lambda(p^2)}{\partial \Lambda} < \infty$\,.
\end{enumerate}
\end{minipage}
\end{equation}
We will show that cutoff functions $K_\Lambda(p^2)$ satisfying these requirements exist in abundance, and can be chosen to take somewhat simple forms.

For illustrative purposes, we find such a $K_\Lambda(p^2)$ explicitly. Recall that
\begin{equation}
  B(x) = \begin{cases} \exp\!\left(1-\frac{1}{1-x^2}\right) & \text{if }\,|x| \leq 1 \\
  0 & \text{if }\,|x| > 1
  \end{cases}
\end{equation}
is a bump function supported on $[-1,1]$ such that $B(1) = B(-1) = 0$.  Moreover, all higher derivatives of $B$ at $x = \pm 1$ equal zero as well.  Let us fix some auxiliary scale $\Lambda_{\text{max}}$, which we take to be some scale larger than our initial value of the UV cutoff $\Lambda_0$.  It will be convenient to specify $K_\Lambda(p^2)$ in three regimes: (i) $\Lambda = \Lambda_{\text{max}}$, (ii) $\Lambda < \Lambda_{\text{max}}$, and (iii) $\Lambda > \Lambda_{\text{max}}$.

At $\Lambda = \Lambda_{\text{max}}$, we fix $K_{\Lambda}(p^2)$ such that this function satisfies conditions 1, 2, and 3 of~\eqref{eq:cutoff-requirements}:
\begin{equation}
\label{E:good-cutoff-1}
    K_{\Lambda = \Lambda_{\text{max}}}(p^2) =  
    \begin{cases}
    1 & p^2 \leq \Lambda_{\text{max}}^2 \\ \\
    (1-\varepsilon)\,B((p^2-\Lambda_{\text{max}}^2)/\varepsilon) + \varepsilon & \Lambda_{\text{max}}^2 \leq p^2 \leq \Lambda^2_{\text{max}}+ \varepsilon \\ \\
    \varepsilon\, e^{-(p^2-(\Lambda^2_{\text{max}}+\varepsilon))} & p^2 > \Lambda^2_{\text{max}} + \varepsilon
    \end{cases}\,.
\end{equation}
We then define $K_{\Lambda}(p^2)$ for $\Lambda < \Lambda_{\text{max}}$ by requiring that the function satisfies conditions 1, 2, and 3 of~\eqref{eq:cutoff-requirements}, is identically $\varepsilon$ from $\Lambda^2+\varepsilon$ up to $\Lambda_{\text{max}}^2 + \varepsilon$, and then agrees with the exponentially decaying tail of $K_{\Lambda= \Lambda_{\text{max}}}(p^2)$ for all larger values of $p^2$:
\begin{equation}
\label{E:good-cutoff-2}
   K_{\Lambda < \Lambda_{\text{max}}}(p^2) = \begin{cases}
    1 & p^2 \leq \Lambda^2 \\ \\
    (1-\varepsilon)\,B((p^2-\Lambda^2)/\varepsilon) + \varepsilon & \Lambda^2 \leq p^2 \leq \Lambda^2+\varepsilon \\ \\
    \varepsilon & \Lambda^2+\varepsilon \leq p^2 \leq \Lambda^2_{\text{max}} + \varepsilon \\ \\
    \varepsilon\, e^{-(p^2-(\Lambda^2_{\text{max}}+\varepsilon))} & p^2 > \Lambda^2_{\text{max}} + \varepsilon
    \end{cases}\,.
\end{equation}
Finally, for $\Lambda > \Lambda_{\text{max}}$\,, we define $K_{\Lambda}(p^2)$ by requiring that this function satisfies conditions 1, 2, and 3 of~\eqref{eq:cutoff-requirements}, but cuts off more sharply in the interval $\Lambda^2 \leq p^2 \leq \Lambda^2+\varepsilon$ such that for $p^2 \geq \Lambda^2+\varepsilon$ the function still agrees identically with the exponentially decaying tail of $K_{\Lambda = \Lambda_{\text{max}}}(p^2)$:
\begin{equation}
\label{E:good-cutoff-3}
       K_{\Lambda > \Lambda_{\text{max}}}(p^2) = \begin{cases}
    1 & p^2 \leq \Lambda^2 \\ \\
    \ (1- \varepsilon\, e^{-\Lambda^2+ \Lambda^2_{\text{max}}})\,B((p^2-\Lambda^2)/\varepsilon\, e^{-\Lambda^2+ \Lambda^2_{\text{max}}})+ \varepsilon\, e^{-\Lambda^2+ \Lambda^2_{\text{max}}}  & \Lambda^2 \leq p^2 \leq \Lambda^2+\varepsilon \\ \\
    \varepsilon\, e^{-(p^2-(\Lambda^2_{\text{max}}+\varepsilon))} & p^2 \geq \Lambda^2+\varepsilon
    \end{cases}\,.
\end{equation}

\begin{figure}
    \centering
    \includegraphics[scale = .5]{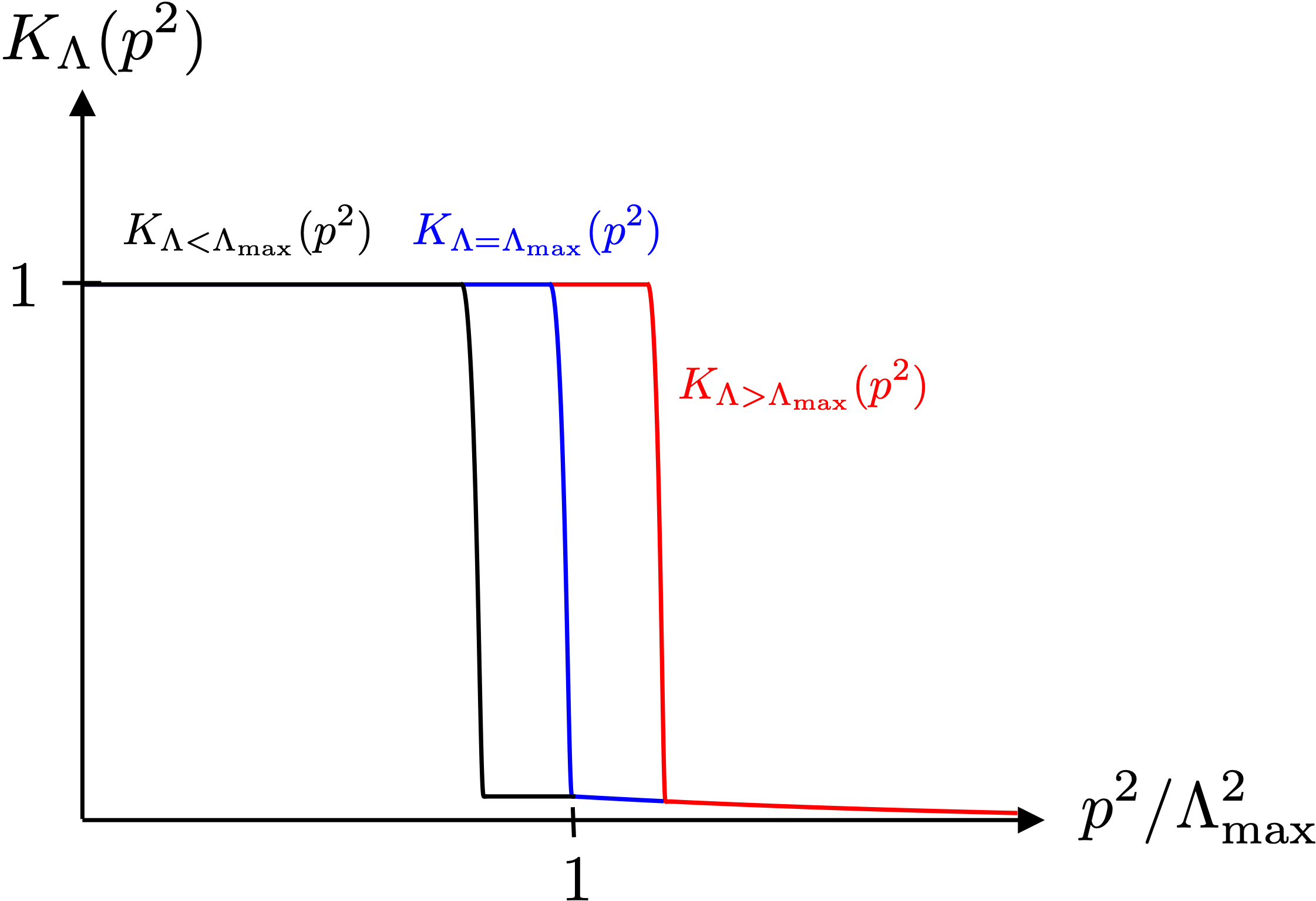}
    \caption{Depiction of the cutoff function $K_\Lambda(p^2)$ given by~\eqref{E:good-cutoff-1},~\eqref{E:good-cutoff-2}, and~\eqref{E:good-cutoff-3}.}
    \label{Fig:soft3}
\end{figure}

The only condition of \eqref{eq:cutoff-requirements} that remains to be verified is the fourth one. However, with our choice of $K_{\Lambda}(p^2)$ specified in~\eqref{E:good-cutoff-1},~\eqref{E:good-cutoff-2}, and~\eqref{E:good-cutoff-3} above, it is clear that for each fixed $\Lambda$, the derivative $\frac{\partial}{\partial \Lambda} K_{\Lambda}(p^2)$ is supported in $\Lambda^2 \leq p^2 \leq \Lambda^2 + \varepsilon$. By the chain rule, the same holds for $\frac{\partial}{\partial \Lambda} \log K_{\Lambda}(p^2)$, rendering this quantity integrable since it is continuous and supported on a compact set. Thus condition 4 of~\eqref{eq:cutoff-requirements} is also satisfied.  A schematic of the cutoff function specified by~\eqref{E:good-cutoff-1},~\eqref{E:good-cutoff-2}, and~\eqref{E:good-cutoff-3} is shown in Figure~\ref{Fig:soft3}.

Recall from our discussion above that the condition~\eqref{E:integral-condition} requires that the factor by which one suppresses large frequencies as one infinitesimally varies $\Lambda$ should decay rapidly with $p^2$. Thus, in the extreme but simple case of the cutoff function $K_\Lambda(p^2)$ specified in~\eqref{E:good-cutoff-1},~\eqref{E:good-cutoff-2}, and~\eqref{E:good-cutoff-3}, large frequencies are not suppressed beyond the fixed initial exponential suppression at frequencies above $\Lambda_{\text{max}}$ (see the last case in each piecewise function definition). One can construct cutoff functions which offer additional suppression of large frequencies by using $\Lambda$-dependent functions more complicated than exponentials.

As an aside, we note that our explicit construction for $K_\Lambda(p^2)$ above is not a differentiable function of $p^2$ at $p^2 = \Lambda^2 + \varepsilon$. However, the RG flows we considered in this paper do not contain $p$-derivatives of $K_{\Lambda}(p^2)$, and so the requirement of $p^2$-differentiability is not necessary.  Nonetheless, by adding a small interpolating region before the last case of each of~\eqref{E:good-cutoff-1},~\eqref{E:good-cutoff-2}, and~\eqref{E:good-cutoff-3}, we can modify $K_\Lambda(p^2)$ so that it is smooth everywhere at the cost of somewhat more complicated formulae.

\bibliography{refs}
\bibliographystyle{JHEP}

\end{document}